\PassOptionsToPackage{dvipsnames}{xcolor}  
\documentclass[reprint,aps,prb,longbibliography,showpacs,amsmath,amssymb,superscriptaddress,floatfix]{revtex4-2}
\setcitestyle{super}
\usepackage{booktabs}
\usepackage{hhline}
\usepackage{amsmath}
\usepackage{amssymb}
\usepackage[version=4]{mhchem}
\usepackage{siunitx}
\DeclareSIUnit\angstrom{\text{\AA}}
\usepackage{svg}
\usepackage{graphicx}
\usepackage{dcolumn}
\usepackage{bm}
\usepackage{placeins}  

\usepackage{hyperref}
\hypersetup{
    unicode=false,
    pdftoolbar=true,
    pdfmenubar=true,
    pdffitwindow=false,
    pdfstartview={FitH},
    colorlinks=true,
    linkcolor=blue,
    citecolor=blue,
    urlcolor=blue,
}
\usepackage{mathptmx}
\usepackage{etoolbox}
\graphicspath{Figures}
\usepackage{multirow}
\usepackage{makecell}  

\pdfoutput=1

\renewcommand{\thetable}{\arabic{table}}  
\usepackage[capitalise]{cleveref}
\usepackage{pifont}

\usepackage[dvipsnames]{xcolor}  
\usepackage[sectionbib]{bibunits}  
\defaultbibliography{SK_zotero}
\defaultbibliographystyle{apsrev4-2}  
\makeatletter
\AtBeginDocument{\let\selectlanguage\@gobble}
\makeatother
\begin{document}
\begin{bibunit}  

\title{Fast and Accurate Equivariant Foundation Models for Atomistic Simulation}
\newcommand\HARVARD{John A. Paulson School of Engineering and Applied Sciences,
Harvard University, Cambridge, MA, USA}

\author{Seán R. Kavanagh}
\email{sk2045@cam.ac.uk}
\affiliation{Yusuf Hamied Department of Chemistry, University of Cambridge, Cambridge, UK}

\author{Chuin Wei Tan}
\affiliation{\HARVARD}

\author{Menghang Wang}
\affiliation{\HARVARD}

\author{Marc L. Descoteaux}
\affiliation{\HARVARD}

\author{Gabriel de Miranda Nascimento}
\affiliation{Department of Materials Science and Engineering, Massachusetts Institute of Technology, Cambridge, MA, USA}

\author{Ulrik Unneberg}
\affiliation{\HARVARD}

\author{Laura Zichi}
\affiliation{\HARVARD}

\author{Francesco Libbi}
\affiliation{\HARVARD}

\author{Norma Rivano}
\affiliation{\HARVARD}


\author{Austin Glover}
\affiliation{Department of Molecular Science and Software Engineering, University of California, Berkeley, CA, USA}

\author{Vivek Bharadwaj}
\affiliation{Department of Molecular Science and Software Engineering, University of California, Berkeley, CA, USA}

\author{Anders Johansson}
\affiliation{Sandia National Laboratories, Albuquerque, NM, USA}

\author{William C. Witt}
\affiliation{\HARVARD}

\author{Albert Musaelian}
\affiliation{Mirian Technologies Inc., Boston, MA, USA}

\author{Boris Kozinsky}
\email{allegro-nequip@g.harvard.edu}
\affiliation{\HARVARD}
\affiliation{Robert Bosch LLC Research and Technology Center, Watertown, MA, USA}

\begin{abstract}
Machine-learned interatomic potentials (MLIPs) have emerged as a transformative tool for computational materials science and chemistry, with universal potentials trained on large and diverse datasets now routinely deployed as `foundation models' for downstream fine-tuning in targeted chemical spaces. Many scientific applications of the resulting models, such as molecular dynamics (MD), require high inference and training speeds as well as accuracy. 
In this work, we examine the limits of equivariant MLIPs, which directly encode physical symmetries in model architectures, to achieve these competing targets -- particularly in the regime of extremely large datasets where data efficiency is less critical.
We show how this trade-off can be addressed, and present a family of foundation potentials in the NequIP and Allegro equivariant MLIP architectures which achieve leading inference speeds and strong scalability as well as excellent accuracies across a range of community benchmarks -- spanning materials discovery, thermal conductivity prediction, and near-equilibrium mechanical and thermodynamic properties.
Accelerations implemented within the NequIP infrastructure now permit training of high-accuracy foundation potentials on ultra-large datasets with dramatically reduced computational cost.
Alongside, we show that efforts to improve model accuracy for materials discovery should focus on dataset diversity and improved, consistent descriptions of transition metal compound energy surfaces.
\end{abstract}

\maketitle

\section{Introduction}
Machine-learned interatomic potentials (MLIPs) have become an indispensable tool for computational materials science and chemistry.\cite{behler_generalized_2007,mroz_crossdisciplinary_2025,mannodi-kanakkithodi_accelerating_2026}
The recent availability of large and diverse datasets of first-principles simulation data (primarily Density Functional Theory (DFT)), such as MPtrj,\cite{deng_chgnet_2023} Alexandria,\cite{schmidt_improving_2024} OMat24,\cite{barroso-luque_open_2024} MatPES,\cite{kaplan_foundational_2025} MAD,\cite{mazitov_petmad_2025} and OMol25\cite{levine_open_2025} has enabled training of MLIPs which are capable of predicting energies and forces over wide chemical spaces with reasonable accuracies -- often termed `universal potentials'. 
Potentials trained on large datasets are being used increasingly as `foundation models', where they provide a pre-trained set of weights and parameters which can then be re-trained, or fine-tuned, on smaller datasets within a target chemical space.\cite{kaur_dataefficient_2025,mosquera-lois_machinelearning_2024,gardner_understanding_2026}
Fine-tuning has been shown to significantly decrease additional training dataset size requirements for downstream application, while yielding comparable predictive accuracies to bespoke models\cite{kaur_dataefficient_2025,kim_dataefficient_2025,gardner_understanding_2026}. While this approach reduces the time to generate training data, the size and inference speed of the resulting fine-tuned model may be a limiting factor in the scientific simulation workflow. 
Moreover, training large universal models on massive datasets has remained a difficult and computationally-expensive task, often requiring weeks, thus impeding architectural and hyperparameter explorations.\\

To address these outstanding issues, in this work, we present a family of equivariant graph neural network foundation models from the NequIP\cite{batzner_e3equivariant_2022} and Allegro\cite{musaelian_learning_2023} model frameworks, trained on the MPtrj,\cite{deng_chgnet_2023} Alexandria,\cite{schmidt_improving_2024} and OMat24\cite{barroso-luque_open_2024} datasets. 
The NequIP\cite{batzner_e3equivariant_2022} framework pioneered the development of interatomic potentials based on equivariant graph neural networks (GNNs), where physical symmetries such as rotation and inversion, as well as translation, are directly incorporated within the model architecture. 
Equivariant models have been shown to improve accuracies, data efficiency and generalisation for MLIPs,\cite{batzner_advancing_2023,batatia_mace_2023,bochkarev_graph_2024}, with several state-of-the-art GNN models based on the NequIP architecture.\cite{merchant_scaling_2023,park_scalable_2024,batatia_design_2025}
Allegro\cite{musaelian_learning_2023} built on the equivariant architecture to employ strictly-local representations without atom-centred message passing, achieving highly-scalable parallel inference on large systems.\cite{kozinsky_scaling_2023}
Notably, Allegro uses atom \emph{pairs} as the central model feature (graph edges), along with iterated tensor products of equivariant representations to model many-body interactions, rather than atoms as in NequIP.\\

This work was facilitated by recent infrastructural developments\cite{tan_highperformance_2026} in the NequIP framework, including train-time graph compilation, distributed data-parallel (DDP) strategies for scalable multi-GPU training and optimised tensor product kernels,\cite{bharadwaj_efficient_2025,noauthor_accelerate_2024} which allow efficient and scalable training on large datasets.
We validate the accuracies of these models using a variety of community benchmarks, including geometry relaxations and materials discovery,\cite{riebesell_framework_2025} predictions of non-equilibrium phonon-related properties,\cite{kaplan_foundational_2025} and thermal conductivities,\cite{pota_thermal_2026} demonstrating excellent performance.
We find that accuracies on current inorganic materials discovery benchmarks are predominantly limited by compositional diversity and the description of potential energy surfaces for transition metal compounds in training datasets.
Finally, we demonstrate the leading inference speeds and excellent scalability of our trained models.

\begin{figure*}[!htbp]
\centering
\includegraphics[width=\linewidth]{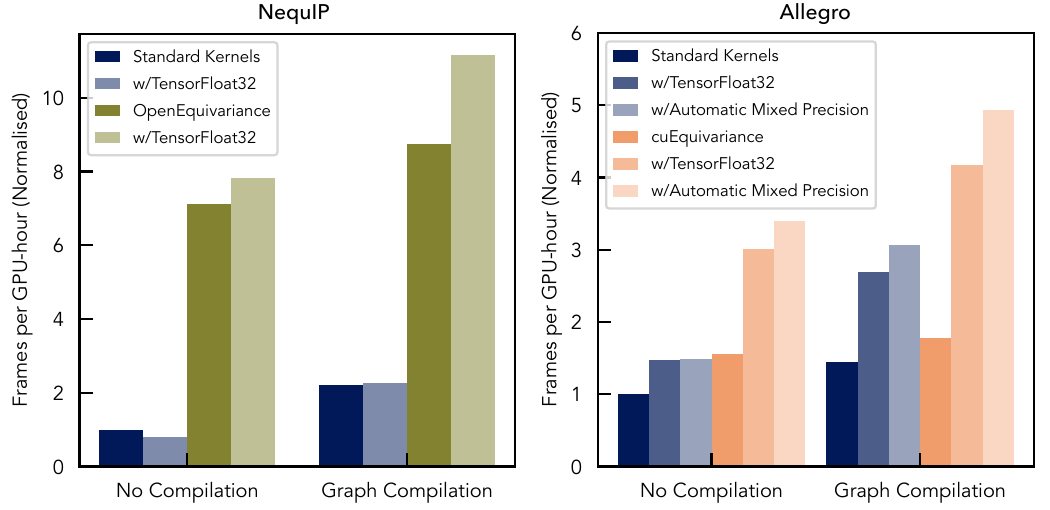}
\caption{\textbf{Accelerated Model Training}. Training speeds for large NequIP and Allegro models with the MPtrj dataset,\cite{deng_chgnet_2023} plotted as the number of data frames (atomic structures with energy and forces) trained on per GPU-hour.
Training speeds are measured without and with graph compilation (\texttt{torch.compile}), using standard tensor product kernels or the recently-implemented \texttt{OpenEquivariance}\cite{bharadwaj_efficient_2025} (for NequIP) and \texttt{cuEquivariance}\cite{noauthor_accelerate_2024} (for Allegro) kernels, and without or with mixed floating-point precision strategies (\texttt{TensorFloat32} or Automatic Mixed Precision\cite{_bfloat16_}).
Training times were recorded using a single node of 4 NVIDIA A100 40 GB GPUs on the NERSC Perlmutter system, with further details given in \cref{sec:methods}.
The same batch sizes were used to standardise comparisons, but we note that mixed-precision and accelerated kernels also significantly reduce GPU memory demand, allowing larger batch sizes and thus second-order train-time accelerations.
}
\label{fig:train_times}
\end{figure*}

\section{Training Accelerations}\label{sec:training_acceleration}
\cref{fig:train_times} shows the acceleration in model \textit{training} provided by recent additions of graph compilation and accelerated tensor-product kernels to the NequIP framework,\cite{tan_highperformance_2026} giving 5-10x cumulative acceleration in model training.
Mixed-precision was also used to accelerate training.
Specifically, double floating-point precision (\texttt{float64}) was used for node embeddings, before casting to single-precision (\texttt{float32}) for tensor products and casting back to double-precision for model read-out, matching the default settings in NequIP and Allegro.
Mixed floating-point precision\cite{_bfloat16_,kharya_nvidia_2020} was then used for intermediate computations (such as matrix multiplications) during model \emph{training}, showing a negligible effect on accuracies while giving up to 1.35x and 3x train-time speedups for NequIP and Allegro respectively, depending on model hyperparameters.
When employing `automatic mixed precision' with the \texttt{bfloat16}\cite{_bfloat16_} format, all \texttt{scatter\_add} operations were reduced in higher precision (\texttt{float32}) to stabilise training -- now implemented by default in the NequIP infrastructure.
The impact of mixed-precision on accuracy depends on data magnitudes, with greater effects on errors for datasets with larger absolute values, such as in the OMol25\cite{levine_open_2025} molecular dataset.
In such cases, mixed-precision can be used for the majority of training epochs, followed by a short continuation with full precision, to achieve similar train-time speed-up without loss in accuracy.\cite{fu_learning_2025}
We note that mixed-precision (including the use of \texttt{TensorFloat32} precision) was disabled for inference tasks in this work, as it was found to introduce noise and slow convergence.\\

In addition to these per-GPU training accelerations, we leveraged the custom Distributed Data Parallel (DDP) training strategy in NequIP\cite{tan_highperformance_2026} to allow distributed multi-GPU training and reduce training times.
Combined, these accelerations allow the training of large foundation potentials on ultra-large datasets of over 100M structures (e.g. OMat24\cite{barroso-luque_open_2024}) with a relatively modest cost of $\sim$100 -- 750 GPU hours (with NVIDIA H100/H200 GPUs).
This dramatically lowers the barrier to training high-accuracy foundation potentials, which we hope will facilitate ready extension to advanced architectures within the community (such as charge-aware models).\cite{cheng_latent_2025,ko_fast_2025}

\begin{table*}[htb]
\def\arraystretch{1.25}
\caption{Key model hyperparameters for NequIP and Allegro MLIPs trained in this work. $l_{\textrm{max}}$ refers to the maximum rotation order of the neural network model features. MLP = Multi-layer perceptron. \\
}  
\centering
\begin{tabular*}{\textwidth}{@{\extracolsep{\fill}}l | c | *{5}{c} r@{}}
\toprule
\textbf{Model} & \textbf{Size} & \textbf{Radial Cutoff (\AA)} & \textbf{$l_{\textrm{max}}$} & \thead{Tensor Features \\ ($l$ = 0, 1... $l_{\textrm{max}}$)} & \thead{Message-Passing \\ Layers} & \thead{MLP Depth} & \textbf{Parameters} \\
\midrule
\multirow{4}{*}{NequIP}
& Small & 4.5 & 1 & 128, 64 & 2 & -- & 0.6 M \\
& Medium & 6 & 2 & 128, 64, 32 & 4 & -- & 3.2 M \\
& Large & 6 & 3 & 128, 64, 32, 32 & 6 & -- & 6.8 M \\
& Extra-Large & 6 & 4 & 320, 96, 64, 32, 32 & 6 & -- & 32.1 M \\
\midrule
\multirow{3}{*}{Allegro} 
& Small & 5 & 2 & 64, 64, 64 & 3 & 1 & 1.4 M \\
& Medium & 6 & 2 & 64, 64, 64 & 5 & 1 & 7.2 M \\
& Large & 7 & 3 & 64, 64, 64, 64 & 5 & 2 & 12.7 M \\
\bottomrule
\end{tabular*}
\label{tab:model_hyps}
\end{table*}

\section{Large Model Training}\label{sec:training}
The key hyperparameters chosen for NequIP and Allegro models trained in this work are shown in \cref{tab:model_hyps}.
A range of models were trained for both architectures, with differing prioritisation of accuracy and speed.
In general, we find that Allegro model training takes approximately twice as many epochs to converge as with NequIP, likely related to its strictly-local / pair-wise representation.
For both architectures, we find a two-stage training to yield optimal results, with a greater weighting of force errors in the loss function for the initial training run until force errors have converged, followed by a short continuation run with 5-10x increased weighting of energy errors in the loss, a reduced learning rate and no mixed-precision.\\

NequIP and Allegro models were trained on four inorganic materials datasets of increasing size and structural diversity, all derived from semi-local DFT calculations. 
The smallest is the DIRECT\cite{qi_robust_2024} dataset of $\sim$186k structures, constructed by DImensionality-Reduced Encoded Clusters with sTratified (DIRECT) sampling of the \texttt{MPF.2021.2.8.All} dataset of Materials Project geometry relaxations (itself a sub-set of compounds and geometries from the MPtrj\cite{deng_chgnet_2023} dataset).
The Materials Project Trajectory (MPtrj) dataset\cite{deng_chgnet_2023} comprises $\sim$1.58M structures sampled from geometry relaxations and static calculations in the  Materials Project database as of September 2022. 
The MPA dataset ($\sim$10.5M structures) combines MPtrj with the Alexandria dataset\cite{schmidt_improving_2024} of equilibrium and near-equilibrium structures, sub-sampled to remove structure prototypes overlapping with the Matbench Discovery\cite{riebesell_framework_2025} materials discovery test set (sAlex)\cite{barroso-luque_open_2024}.
The largest, OAM ($\sim$113M structures), incorporates the OMat24 dataset\cite{barroso-luque_open_2024} of non-equilibrium geometries from \emph{ab initio} molecular dynamics and relaxation trajectories of rattled structures from the Alexandria\cite{schmidt_improving_2024} dataset, along with MPA.
When training with the OAM dataset, we first train with just the OMat24 dataset\cite{barroso-luque_open_2024}, before fine-tuning on the MPA dataset to ensure compatibility with Materials Project\cite{jain_materials_2013} calculation settings (as required for the Matbench Discovery\cite{riebesell_framework_2025} benchmark tasks).
In this manuscript, we use the \texttt{\{architecture\}-\{dataset\}-\{model size\}} syntax to refer to the various trained models, where \texttt{architecture} is either NequIP or Allegro, \texttt{dataset} is DIRECT, MP (for MPtrj), MPA or OAM, and \texttt{model size} is S, M, L or XL for small, medium, large or extra-large, respectively. 

\section{Benchmark Performance}\label{sec:benchmarks}
\begin{table*}[htbp]
\def\arraystretch{1.25}
\caption{Performance metrics for the NequIP and Allegro models, trained on the MP and OAM datasets, with accuracy benchmarks from the Matbench Discovery\cite{riebesell_framework_2025} and MatCalc\cite{kaplan_foundational_2025} test suites. This includes the mean absolute error (MAE) in convex hull energy predictions for unseen compounds ($E_{\textrm{WBM, MAE}}$), the root-mean-square deviation (RMSD) in predicted geometries of unseen compounds, dimensionless Symmetric Relative Mean Error in thermal conductivities for 103 binary compounds ($\kappa_{\textrm{SRME}}$), MAE in constant-volume heat capacity ($C_{v\textrm{, MAE}}$), and mean ratio of predicted to reference force magnitudes ($f/f_{\text{DFT}}$).
Model hyperparameters, training datasets and naming syntax are described in \cref{sec:training}, with additional details of benchmark evaluation parameters in \cref{sec:methods}. The NequIP-OAM-XL model yields the best accuracies across each benchmark.
} 
\label{tab:benchmarks}
\centering
\begin{tabular*}{\textwidth}{@{\extracolsep{\fill}}l | *{5}{c}@{}}
\toprule
Model &  
\thead{$E_{\textrm{WBM, MAE}}$\cite{riebesell_framework_2025}\\(meV/atom)} & 
\thead{\ \ RMSD\cite{riebesell_framework_2025}\\(\AA)} & 
\thead{$\kappa_{\textrm{SRME}}$\cite{pota_thermal_2026}} &
\thead{\ \ $C_{v\textrm{, MAE}}$\cite{kaplan_foundational_2025}\\(J~mol$^{-1}$~K$^{-1}$)} & 
\thead{\ \ \ \ $f/f_{\text{DFT}}$\cite{kaplan_foundational_2025,deng_systematic_2025}\\(\%)} \\
\midrule
\midrule
Allegro-MP-L  & 44.3 & 0.083 & 0.486 & 6.95 & 92.8 \\
NequIP-MP-L  & 42.7 & 0.087 & 0.368 & 7.51 & 96.7 \\
\midrule
\midrule
Allegro-OAM-S & 32.9 & 0.072 & 0.673 & 7.33 & 95.3 \\
Allegro-OAM-M & 24.7 & 0.067 & 0.437 & 5.86 & 96.0 \\
Allegro-OAM-L & 21.4 & 0.063 & 0.313 & 4.38 & 96.9 \\
NequIP-OAM-S & 45.9 & 0.094 & 0.855 & 13.7 & 86.1 \\
NequIP-OAM-M & 24.5 & 0.067 & 0.254 & 5.60 & 95.0 \\
NequIP-OAM-L & 21.5 & 0.063 & 0.163 & 4.49 & 97.3 \\
NequIP-OAM-XL & 19.7 & 0.060 & 0.129 & 4.02 & 97.9 \\
\bottomrule
\end{tabular*}
\end{table*}

To demonstrate the accuracy of these foundation potentials, we evaluate their performance on a set of established community benchmark tasks,\cite{riebesell_framework_2025,kaplan_foundational_2025,pota_thermal_2026} with results given in \cref{tab:benchmarks}.
Briefly, this includes:
\begin{itemize}
    \item \textbf{Materials Discovery:} 
    One of the most common workflows in computational materials science is the prediction of the ground-state geometry and energy of a given atomic configuration, i.e. the minimum of the potential energy surface (PES), using gradient relaxation.
    The \texttt{matbench-discovery}\cite{riebesell_framework_2025} structural relaxation benchmark assesses model accuracies for this task, by considering the error in predicted convex hull energies ($E_{\textrm{WBM, MAE}}$) and atomic structures (RMSD) for the WBM un-seen test set of $\sim250,000$ hypothetical inorganic solids.\cite{wang_predicting_2021}
    This test set was originally generated through ionic substitution of chemically-similar species in known materials, followed by geometry relaxations using Density Functional Theory (DFT) to predict thermodynamic stabilities.\cite{wang_predicting_2021}

    \item \textbf{Phonon and Thermal Properties:}
    Many physical attributes of materials are governed by their dynamic -- rather than static -- properties, as dictated by off-equilibrium atomic forces; i.e. PES gradients.
    The accuracy of interatomic potentials for modelling such dynamic properties can be evaluated by considering phonon band structures (representing atomic vibrations) and derived properties, which probe successively higher-order derivatives of the PES.
    At the harmonic level, Loew et al.\cite{loew_universal_2025} recently introduced a large-scale benchmark of phonon band structures based on $\sim$10,000 non-magnetic semiconductors from the MDR phonon database,\cite{_mdr_} which allows us to benchmark predicted maximum phonon frequencies ($\omega_{\textrm{max}}$) -- a sensitive probe of PES curvature, and thus phonon over-softening -- along with Helmholtz free energies, vibrational entropies and heat capacities against a large set of DFT reference values.
    We complement this with a direct, mode-resolved evaluation of phonon frequency errors on the 26,234 Materials Project\cite{jain_materials_2013} structures for which reference DFT phonon data is available, as discussed in \cref{SI:Thermal_Properties_Error_Analysis}.
    Going beyond the harmonic approximation, the benchmark from Póta et al.\cite{pota_thermal_2026} provides a more stringent test, using predicted 2nd- and 3rd-order atomic force constants to compute thermal conductivities for 103 binary compounds, and comparing to DFT reference values.
    This benchmark provides a robust test of off-equilibrium PES accuracy and smoothness (key to stability in molecular dynamics simulations, for instance), and revealed issues with over-fitting to equilibrium properties and force noise in some previous generations of foundation potentials.\cite{riebesell_framework_2025}
    Given that thermal conductivities vary over several orders of magnitude, and to deconvolute the effect of fortuitous error cancellations in phonon properties, the authors introduce a unit-less Symmetric Relative Mean Error ($\kappa_{\textrm{SRME}}$) metric based on single-phonon conductivities for quantifying performance on this benchmark.

    \item \textbf{Bulk Mechanical Properties}  
    Beyond microscopic phonons and related properties, the dynamic response of solids to structural perturbations can be quantified through derived macroscopic properties such as heat capacity ($C_\textrm{V}$), bulk modulus ($K$) and shear modulus ($G$). 
    Kaplan et al.\cite{kaplan_foundational_2025} recently introduced benchmarks for these near-equilibrium properties, which highlighted the importance of quality data to achieve reliable and generalisable MLIPs.
    These benchmarks additionally include a test of force magnitudes on high energy states ($\% |\mathbf{F}_i|$), to quantify PES and phonon over-softening due to biased sampling of near-equilibrium structures in training datasets.\cite{deng_systematic_2025}

\end{itemize}

\begin{figure*}[htbp]
\centering
\includegraphics[width=\linewidth]{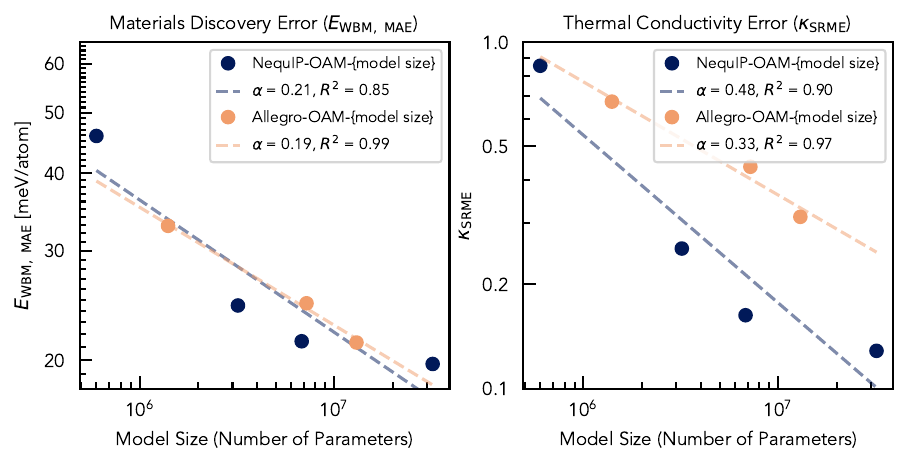}
\caption{\textbf{Model Size Learning Curve}. 
Test errors for NequIP and Allegro models trained on the OAM dataset (NequIP/Allegro-OAM-\texttt{\{model size\}}) and evaluated on the Matbench Discovery\cite{riebesell_framework_2025} benchmark suite, as a function of model size (\cref{tab:model_hyps,tab:benchmarks}).
This includes the mean absolute error (MAE) in convex hull energy predictions for unseen compounds ($E_{\textrm{WBM, MAE}}$, left) and the dimensionless Symmetric Relative Mean Error in thermal conductivities for 103 binary compounds ($\kappa_{\textrm{SRME}}$, right).
Dashed lines show least-squares power law fits to the data ($L = a\cdot x^{-\alpha}$), with the fitted scaling exponent $\alpha$ and coefficient of determination $R^2$ given in the figure legends. 
Model hyperparameters, training datasets and naming syntax are described in \cref{sec:training}.
}
\label{fig:benchmark_by_model_size}
\end{figure*}

\begin{figure*}[htbp]
\centering
\includegraphics[width=\linewidth]{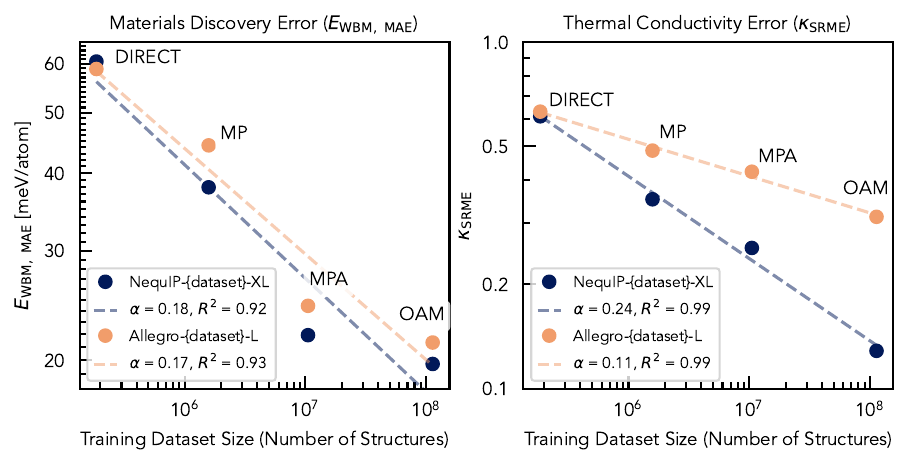}
\caption{\textbf{Training Data Learning Curve}. 
Test errors for the largest NequIP and Allegro models trained in this work (NequIP-\texttt{\{dataset\}}-XL and Allegro-\texttt{\{dataset\}}-L) and evaluated on the Matbench Discovery\cite{riebesell_framework_2025} benchmark suite, as a function of training dataset size. 
This includes the mean absolute error (MAE) in convex hull energy predictions for unseen compounds ($E_{\textrm{WBM, MAE}}$, left) and the dimensionless Symmetric Relative Mean Error in thermal conductivities for 103 binary compounds ($\kappa_{\textrm{SRME}}$, right).
Dashed lines show least-squares power law fits to the data ($L = a\cdot x^{-\alpha}$), with the fitted scaling exponent $\alpha$ and coefficient of determination $R^2$ given in the figure legends. 
Model hyperparameters, training datasets and naming syntax are described in \cref{sec:training}.
}
\label{fig:benchmark_by_dataset_size}
\end{figure*}

We find excellent performance across these benchmarks for the NequIP and Allegro models trained in this work, on par with the best performing models reported to date.\cite{riebesell_framework_2025}
For the most part, we witness the expected trends in benchmark performance with continuously improving accuracies as we scale to increasingly larger model and training dataset sizes (\cref{fig:benchmark_by_model_size,fig:benchmark_by_dataset_size}).
For both the materials discovery and thermal conductivity accuracy benchmarks, we see a consistent power-law relation (near-linear trend on the log-log scale) with model size for the small, medium and large models for both NequIP and Allegro (\cref{fig:benchmark_by_model_size}) -- further illustrated in \cref{sifig:benchmark_by_model_size_3_data_points} by only fitting to these data-points. 
The NequIP-OAM-XL model deviates from the power law trend for both accuracy benchmarks, as the accuracy begins to saturate at these model sizes. 
Moreover, different hyperparameter choices (\cref{tab:model_hyps}) have varying effects on the total number of parameters (which often does not correlate well with true model capacity), and thus ideal power-law scaling is not necessarily expected for the model-size learning curve.\\

For the training data-size learning curves we see mostly similar trends, with ideal power law scaling for the thermal conductivity benchmark ($R^2 = 0.99$), and near-ideal scaling for materials discovery ($R^2 \sim 0.93$), for both NequIP and Allegro.
Imperfect scaling for the materials discovery task here stems from the significant under-performance when scaling from the MPA to OAM datasets -- even more apparent when omitting the OAM datapoint from the fit (\cref{sifig:benchmark_by_dataset_size_3_data_points}) -- despite ideal scaling for the thermal conductivity benchmark. \\

\subsection{Materials Discovery Benchmark Performance}\label{subsec:Materials_Discovery_Benchmark}
\begin{figure*}[!tp]
\centering
\includegraphics[width=\linewidth]{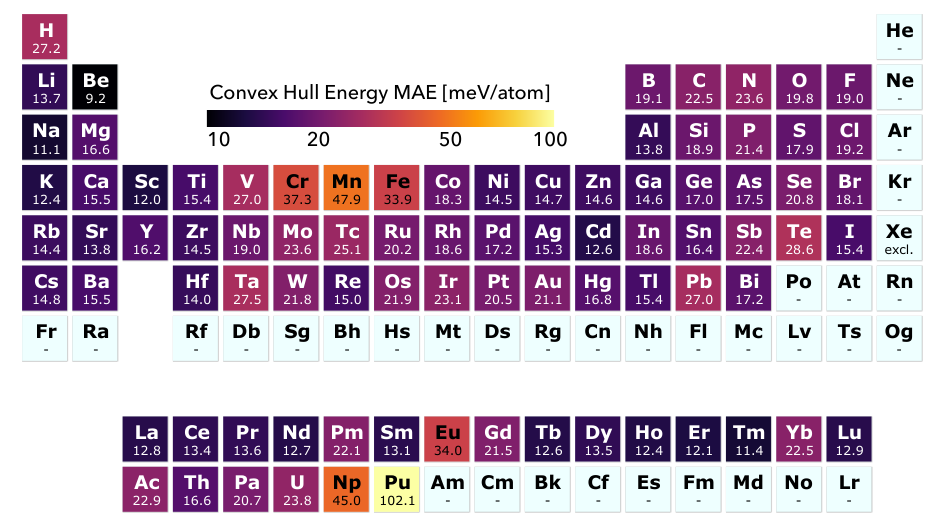}
\caption{\textbf{Per-Element Errors for Materials Discovery}. 
Distribution of element-wise convex hull energy mean absolute error (MAE) across the periodic table for the \texttt{matbench-discovery}\cite{riebesell_framework_2025} materials discovery benchmark, using the NequIP-OAM-XL model.
Errors are shown in meV/atom, using a logarithmic colourbar scale.
The largest element-wise errors are seen for compounds containing multi-valent first-row transition metals (V, Cr, Mn, Fe), low-prevalence actinides (Np, Pu), along with Eu, Ta, Te, Pb and H. 
}

\label{fig:per-element_errors}
\end{figure*}

The lack of a significant improvement in materials discovery accuracy, when expanding from the $\sim$10.5 million structures MPA dataset to the $\sim$113 million structures OAM dataset, can be explained by considering the distribution of training data and materials discovery errors over the periodic table.
In \cref{fig:per-element_errors}, we plot the element-wise materials discovery benchmark errors for the NequIP-OAM-XL model, obtained by averaging the convex hull energy errors across all phases in the test set and weighting by elemental fraction in each phase composition.
From this elemental error distribution, we see that multi-valent transition metals -- particularly the first-row transition metals V, Cr, Mn and Fe -- have consistently higher errors than the overall average; from $\sim$27 -- \SI{48}{meV/atom}, compared to \SI{19.7}{meV/atom} over all elements/compositions.\\

Potential energy surfaces of compounds containing transition metals are notoriously difficult to model, both for metals and insulators.\cite{owen_complexity_2024}
Complex multi-valent and correlated behaviour requires care in quantum training data generation. 
Localised orbitals and highly-directional many-body interactions often require high levels of radial and angular resolution for accurate modelling.\cite{owen_complexity_2024}
Also, universal MLIPs are tasked to learn a single ground-state potential energy surface as a function of only atomic positions and types for a large variety of electronic structure configurations (spin/oxidation states).
Recent work\cite{warford_better_2026} has shown that inconsistencies in the application of Hubbard U corrections in training data propagate into systematic errors in the resulting models.
Hubbard U corrections are selectively applied for oxide and fluoride compounds containing V, Cr, Mn, Fe, Co, Ni, Mo or W, in the inorganic training datasets used in this work (OMat24\cite{barroso-luque_open_2024} Alexandria\cite{schmidt_improving_2024} and MPtrj\cite{deng_chgnet_2023}), but not in metallic or non-oxide/fluoride compositions with these elements, following the Materials Project\cite{jain_materials_2013} default settings. 
We thus ascribe the larger materials discovery errors for (first-row) transition metal compounds to the innate complexity of many-body transition metal bonding and inconsistencies in the application of $+U$ electron correlation corrections, which effectively provide an inconsistent training target.
Np and Pu also stand out as clear high-error outliers in \cref{fig:per-element_errors}, likely the result of low prevalence in the training datasets and again complex multi-valent and highly-correlated electronic behaviour.\\

Excluding the first-row transition metals reduces the materials discovery convex hull energy MAE from 19.7 to \SI{16.3}{meV/atom}, while also omitting Np and Pu compounds further reduces the MAE to \SI{13.9}{meV/atom}, indicating that these outliers are dominating the overall benchmark metrics. 
This analysis helps explain the poor improvement in materials discovery benchmark accuracies when expanding the training dataset from MPA to include OMat24\cite{barroso-luque_open_2024} (i.e. OAM), despite significant improvements in thermal conductivity benchmarks.
The large OMat24 dataset\cite{barroso-luque_open_2024} comprises off-equilibrium structures generated via rattling and molecular dynamics of configurations already present in the Alexandria (MPA) dataset,\cite{schmidt_improving_2024} and so provides useful additional information on energy surface gradients (and thus phonon-related properties) but offering no additional chemical diversity.
Moreover, the same $+U$ correction scheme as the Alexandria\cite{schmidt_improving_2024} and MPtrj\cite{deng_chgnet_2023} datasets was employed, to retain consistency with the Materials Project\cite{jain_materials_2013} computational settings.
Combined, these issues result in minimal improvement in materials discovery benchmark accuracies with the larger OMat24 dataset.\cite{barroso-luque_open_2024}
We note that similar distributions in element-wise relative errors are obtained for other foundation potentials trained on the same datasets and evaluated on the same benchmark (\cref{SI:Other_Model_Element_Wise_Errors}), reflecting that these are general trends related to the data and mostly independent of specific model architectures and hyperparameter choices.
\\

Taken together, this analysis suggests some key directions to improve foundation potential accuracies going forward.
Firstly, avoiding systematic errors from the propagation of inconsistent computational settings / application of correction schemes is crucial.
This could be achieved by ensuring fully-consistent computational parameters across all training data, but is extremely restrictive and would require a large portion of the available training data to be excluded or re-calculated.
Alternatively, appropriately treating these distinct computational setups (i.e. energy surfaces) as separate labels in multi-fidelity frameworks -- such as multi-head architectures -- offers a promising route to retaining useful chemical information while avoiding systematic errors from inconsistent energy surfaces.
Secondly, chemical diversity is evidently a crucial ingredient in training datasets for materials discovery efforts, being a key limiting factor in predictive accuracies here. 
Future efforts in dataset building (for the goal of materials discovery) could focus on expanding both chemical and structural diversity, going beyond ionic substitution and likely requiring structure-searching methods or generative models.
Finally, the diverse range of complexity in atomic environments and interactions poses real challenges.
Multi-valent elements with highly-directional $d$ or $f$ orbitals require high radial and angular resolution in accurate models, while elements with simpler bonding interactions (e.g. alkali metals) have much lower requirements.
Simply raising the radial and angular resolution (i.e. model capacity) across the board results in major inefficiencies due to these diverse requirements -- though becomes increasingly feasible with software accelerations and growth in computing power.
Future architectural developments which allow models to smoothly integrate multi-fidelity representations, based on the central element or other atomic environment features, could allow optimal combinations of accuracy and efficiency.

\subsection{Thermal Properties Benchmark Performance}\label{subsec:Thermal_Properties_Benchmark}
\begin{table*}[tp]
    \caption{MatCalc benchmark results; including structural similarity to DFT references after geometry relaxation ($d_{\textrm{MAE}}$, unit-less), formation energies per atom ($E^f_{\textrm{MAE}}$), bulk modulus ($K_{\textrm{MAE}}$), shear modulus ($G_{\textrm{MAE}}$), constant-volume heat capacity ($C_{v\textrm{, MAE}}$) and off-equilibrium force magnitudes ($f/f_{\text{DFT}}$, unit-less).
    OAM refers to the protocol of initial model training on the OMat24 dataset,\cite{barroso-luque_open_2024} prior to fine-tuning on the sub-sampled Alexandria\cite{schmidt_improving_2024} and MPtrj\cite{deng_chgnet_2023} datasets. The MatterSim dataset contains Alexandria,\cite{schmidt_improving_2024} MPtrj,\cite{deng_chgnet_2023} and closed-source synthetic data. The UMA dataset contains the OC20++,\cite{tran_open_2023} OMat24,\cite{barroso-luque_open_2024} OMol25,\cite{levine_open_2025} ODAC25,\cite{sriram_open_2025} and OMC25\cite{gharakhanyan_open_2025} datasets.
    Allegro-MatPES is an Allegro model trained from scratch on the MatPES\cite{kaplan_foundational_2025} dataset, with $l_{\textrm{max}}=3$ and 64 tensor features.
    The lowest errors are shown in bold for each property.
    }
    \centering
    \setlength{\tabcolsep}{7pt}  
    \begin{tabular}{llcccccc}
        \toprule
        Dataset & Model & $d_{\textrm{MAE}}$ & $E^f_{\textrm{MAE}}$ (meV/atom) & $K_{\textrm{MAE}}$ (GPa) & $G_{\textrm{MAE}}$ (GPa) & $C_{v\textrm{, MAE}}$ (J~mol$^{-1}$~K$^{-1}$) & $f/f_{\text{DFT}}$ (\%)
        \\  
        \midrule
        MatPES & CHGNet\cite{deng_chgnet_2023} & 0.296 & 95.9 & 23.7 & 20.6 & 23.0 & 91.4 \\
        MatPES & TensorNet\cite{simeon_tensornet_2023} & 0.252 & 90.2 & 18.0 & 14.8 & 13.7 & 93.0 \\
        MatPES & Allegro-MatPES & \textbf{0.244} & 90.6 & 18.6 & 14.3 & 10.8 & 91.7 \\
        \addlinespace
        OAM & NequIP-OAM-S & 0.409 & 68.9 & 24.3 & 19.3 & 13.7 & 86.1 \\
        OAM & NequIP-OAM-M & 0.300 & 31.4 & 16.0 & 14.5 & 5.60 & 95.0 \\
        OAM & NequIP-OAM-L & 0.281 & 26.6 & 14.9 & 14.9 & 4.49 & 97.3 \\
        OAM & NequIP-OAM-XL & 0.279 & 25.4 & \textbf{12.7} & \textbf{14.1} & 4.02 & 97.9 \\
        \addlinespace
        OAM & Allegro-OAM-S & 0.309 & 39.3 & 17.4 & \textbf{14.1} & 7.33 & 95.3  \\
        OAM & Allegro-OAM-M & 0.290 & 31.2 & 14.4 & 15.7 & 5.86 & 96.0 \\
        OAM & Allegro-OAM-L & 0.287 & \textbf{24.7} & \textbf{12.7} & 14.7 & 4.38 & 96.9 \\
        \addlinespace
        MPTrj & NequIP-MP-L & 0.377 & 52.0 & 16.2 & 17.6 & 7.51 & 96.7 \\
        \addlinespace
        MPTrj & Allegro-MP-L & 0.360 & 51.7 & 14.0 & 15.0 & 6.95 & 92.8 \\
        \addlinespace
        MatterSim & MatterSim-5M\cite{yang_mattersim_2024} & 0.322 & 40.5 & 16.9 & 15.8 & 5.46 & 97.3 \\
        UMA & UMA-s-1p2\cite{wood_uma_2026} & 0.267 & 25.0 & 14.9 & 21.5 & \textbf{3.60} & \textbf{98.6} \\
        \bottomrule
    \end{tabular}
    \label{tab:matcalc}
\end{table*}

As discussed above, thermal property benchmarks provide robust validation of energy surface smoothness and gradients in the trained models.
In contrast to the materials discovery benchmark, we see a more consistent improvement in thermal conductivity prediction accuracy ($\kappa_{\textrm{SRME}}$) as we increase both the model and (in particular) dataset sizes, as shown in \cref{fig:benchmark_by_model_size,fig:benchmark_by_dataset_size}.
Similar trends are seen with the MDR and Materials Project phonon benchmarks shown in \cref{SI:Thermal_Properties_Error_Analysis} -- where excellent accuracies are again demonstrated by the NequIP-OAM-XL model.
This continuous ideal scaling in thermal property accuracy with training data is expected, given that each successively larger dataset comprises further off-equilibrium geometries, resulting in increasingly accurate models of energy surface gradients.
This trend reiterates the importance of non-equilibrium geometries for training robust and versatile MLIPs.
Future dataset construction efforts targeting non-equilibrium properties (such as finite-temperature and transport phenomena) should continue to prioritise these configurations, ideally being combined with approaches that maximise structural diversity to boost data efficiency.\cite{kaplan_foundational_2025,mazitov_massive_2025}
\\

We note that while the materials discovery accuracies and trends are broadly similar between NequIP and Allegro, the Allegro $\kappa_{\textrm{SRME}}$ values are consistently larger than for NequIP.
In particular, the learning rates for $\kappa_{\textrm{SRME}}$ with Allegro are significantly lower than NequIP, with $\alpha$ = 0.33 and 0.11 for the model and data-size learning curves, respectively, compared to $\alpha$ = 0.48 and 0.24 for NequIP (\cref{fig:benchmark_by_model_size,fig:benchmark_by_dataset_size}).
The origins of this under-performance are not immediately clear, as validation errors for forces -- which we find to most closely correlate with $\kappa_{\textrm{SRME}}$ -- are on par with, if not slightly better than, those of the NequIP models.
We do find that Allegro $\kappa_{\textrm{SRME}}$ values are more sensitive to the atomic displacement distance used in the underlying finite-difference phonon calculations, which has been associated with floating-point precision effects before,\cite{fu_learning_2025,rhodes_orbv3_2025} but increasing the displacement magnitude does not fully close the gap to NequIP.
Decomposing the error by phonon frequency, we see that while the overall $\kappa_{\textrm{SRME}}$ is dominated by the low-energy acoustic range (0--1~THz) as expected, Allegro exhibits consistently larger errors in phonon frequencies than NequIP across the full spectrum (\cref{SI:Thermal_Properties_Error_Analysis}).
Evaluating our set of trained models on the 26,234 Materials Project\cite{jain_materials_2013} structures for which reference DFT phonon data is available, we find similar trends in the relative accuracies of the models -- though with lower average errors than the 103 binary materials in the $\kappa_{\textrm{SRME}}$ test set.
\\

One natural hypothesis is that this small but consistent higher error in phonon properties could reflect the strictly-local architecture of Allegro, which limits its effective receptive field to $2 \times R_{\max}$ (as opposed to $2 \times n_{\mathrm{layers}} \times R_{\max}$ for NequIP and other message-passing GNN architectures, with $R_{\max}$ being the model radial cutoff), neglecting longer-range contributions to the interatomic force constants.
To test this hypothesis, in \cref{sifig:IFC_range} we plot the achievable $\kappa_{\textrm{SRME}}$ accuracy as a function of interaction range, where interatomic force constants (IFCs) are re-fitted on reference DFT displacement-force data of atomic pairs up to a given radial cutoff.
This analysis shows that indeed at the receptive field ranges of the Allegro OAM models, non-negligible errors in phonon properties are introduced by this truncated interaction range. 
Minimum $\kappa_{\textrm{SRME}}$ errors of $\sim0.25 - 0.1$ are found for the 10 -- 14 \AA{} receptive fields of the small to large Allegro OAM models, mostly explaining the offset in $\kappa_{\textrm{SRME}}$ accuracies between Allegro and NequIP OAM models in \cref{fig:benchmark_by_model_size}.
We do note, however, that despite the error introduced by the limited interaction ranges, the Allegro $\kappa_{\textrm{SRME}}$ values are still comparable to other leading model architectures trained on the same datasets.\cite{riebesell_framework_2025}
\\

\subsection{MatCalc Benchmarks}
We additionally evaluate our foundation potentials on the MatCalc benchmark suite,\cite{kaplan_foundational_2025} which tests both near-equilibrium materials properties (structural similarity to DFT references after geometry relaxation ($d_{\textrm{MAE}}$), formation energies ($E^f_{\textrm{MAE}}$), bulk modulus $K$, shear modulus $G$ and constant-volume heat capacity $C_\textrm{V}$) and off-equilibrium PES fidelity (via the ratio of predicted to reference (DFT) force magnitudes on high-energy perturbed structures, $f/f_{\textrm{DFT}}$).
The results are shown in \cref{tab:matcalc}, alongside reported values for other foundation potentials in the literature.
For comparison, we also train an Allegro model from scratch on the MatPES\cite{kaplan_foundational_2025} dataset, using $l_{\textrm{max}}$ = 3, 64 scalar and tensor features, and 3 layers with a width of 128.
The NequIP-OAM-XL model achieves excellent accuracy across near-equilibrium benchmarks, achieving low mean absolute errors (MAEs) in particular for derived macroscopic properties such as the bulk modulus ($K_{\textrm{MAE}}$ = 12.7 GPa), shear modulus ($G_{\textrm{MAE}}$ = 14.1 GPa) and constant-volume heat capacity ($C_{\textrm{v, MAE}}$ =  4.02~J~mol$^{-1}$~K$^{-1}$).
\\

Moreover, we find that over-softening, where MLIPs trained primarily on near-equilibrium structures tend to systematically under-estimate forces in high-energy/force regions,\cite{deng_systematic_2025} is minimal for our (extra-)large models, with an $f/f_{\textrm{DFT}}$ ratio of \SI{98}{\%}.
From \cref{tab:matcalc}, we do see, however, that with decreased model size, over-softening becomes significant, with an $f/f_{\textrm{DFT}}$ ratio of \SI{86}{\%} for the NequIP-OAM-S model.
Surveying the $f/f_{\textrm{DFT}}$ ratios in \cref{tab:benchmarks,tab:matcalc} for NequIP/Allegro models of different model sizes and training datasets (MP vs OAM), we see that both model size and dataset influence over-softening.
Larger models and datasets with greater off-equilibrium configurations (i.e. Alexandria\cite{schmidt_improving_2024} and OMat24\cite{barroso-luque_open_2024}) yield reduced over-softening, but with model size being the surprisingly larger factor here -- though we note that this is of course dependent on the specific datasets and model hyperparameters being compared.

\subsection{Accuracy-Speed Trade-off}\label{subsec:Accuracy_and_Speed}

\begin{figure*}[htbp]
\centering
\includegraphics[width=\linewidth]{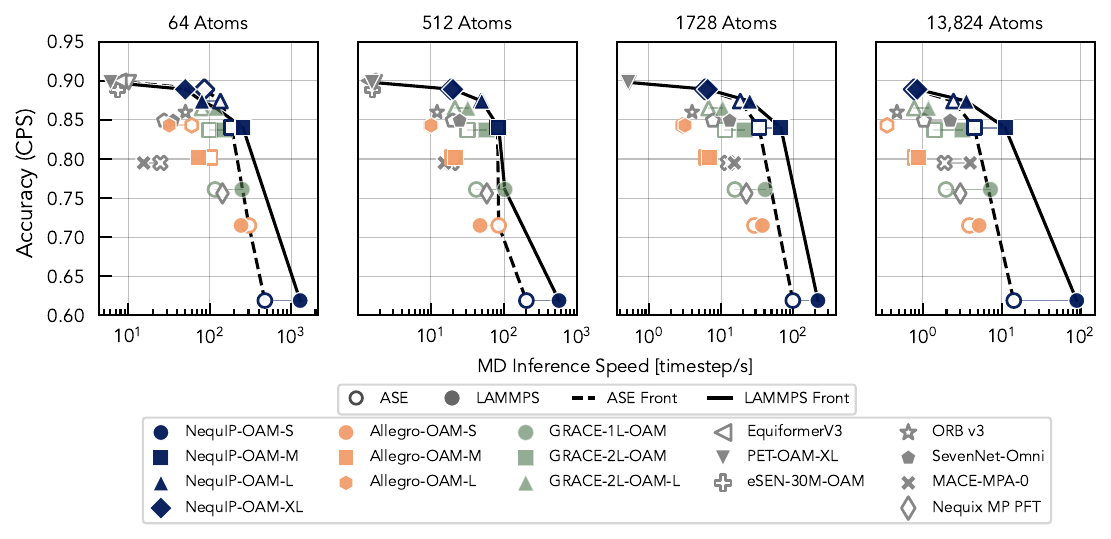}
\caption{\textbf{Accuracy-Speed Trade-off}. \texttt{matbench-discovery}\cite{riebesell_framework_2025} `Combined Performance Score' (CPS) plotted against Molecular Dynamics (MD) inference speed for Si diamond systems ranging from 64 to 13{,}824 atoms, using a single NVIDIA A100-SXM4 (80\,GB) GPU.
MD simulations were performed with both the Atomic Simulation Environment (ASE)\cite{larsen_atomic_2017} (hollow markers) and \texttt{LAMMPS}\cite{thompson_lammps_2022} (filled markers) -- for models with \texttt{LAMMPS} interfaces -- using the publicly-documented setup and available accelerators for each model.
In each panel, the Pareto front of accuracy versus inference speed is shown as dashed and solid black lines for ASE and \texttt{LAMMPS}.
In the current version, CPS averages the materials discovery, thermal conductivity and structural similarity \texttt{matbench-discovery}\cite{riebesell_framework_2025} benchmark accuracies (\cref{tab:benchmarks}) with a relative weighting of 5:4:1.
Models missing for larger system sizes crashed with out-of-memory errors.
}
\label{fig:pareto}
\end{figure*}

In \cref{fig:pareto}, we plot the accuracies of our trained NequIP and Allegro models, as given by the current \texttt{matbench-discovery}\cite{riebesell_framework_2025} `combined performance score' (CPS), against their inference speeds in molecular dynamics production runs, alongside corresponding values for other foundation potentials in the literature. 
In its current form, CPS averages the materials discovery, thermal conductivity, and structural similarity (RMSD) accuracies with a relative weighting of 5:4:1.\cite{riebesell_framework_2025}
Here we measure inference speeds with both the Atomic Simulation Environment (ASE)\cite{larsen_atomic_2017} -- to allow direct comparability with a range of other community models, many of which lack \texttt{LAMMPS} interfaces -- and the \texttt{LAMMPS}\cite{thompson_lammps_2022} MD engine -- where possible.
As expected, we see that \texttt{LAMMPS} allows greater inference speeds for larger systems and smaller models (along with supporting multi-GPU simulations; \cref{sec:scaling}), while minimal differences with ASE are seen for larger models and smaller systems where pure GPU model inference dominates.
For medium-sized models with small system sizes (64 -- 512 atoms), ASE can give faster inference than LAMMPS for NequIP and Allegro, by supporting the combination of (1) ahead-of-time (AOTInductor) compilation via \texttt{nequip-compile} and (2) accelerated tensor-product kernels -- while the \texttt{LAMMPS} interfaces currently only support one or other of these accelerations, but not both.\\

We see that our NequIP models define the accuracy--speed Pareto front across a wide range of model sizes and inference speeds.
For instance, despite the same combined performance score as the eSEN-30M-OAM\cite{fu_learning_2025} model, the NequIP-OAM-XL model is approximately an order of magnitude faster at inference time.
The large, medium and small models offer further substantive speedups at accuracies competitive with leading published models.
We attribute this favourable accuracy-speed trade-off of the NequIP models to a combination of accelerated tensor-product kernels, which minimise the compute cost of equivariant operations,\cite{bharadwaj_efficient_2025,noauthor_accelerate_2024} and judicious model hyperparameter choices. 
We note that other community benchmarks where we have uploaded the NequIP/Allegro OAM models, such as the \texttt{CatBench}\cite{moon_catbench_2025} suite, also show these models exhibiting leading inference speeds and good accuracies.\cite{chiang_mlip_2025}
The full set of single-GPU \texttt{LAMMPS} MD inference speed benchmarks for the subset of models with \texttt{LAMMPS} interfaces are shown in \cref{sifig:lammps_speeds}, where the NequIP and Allegro models, particularly through the Kokkos-accelerated ML-IAP\cite{johansson_lammpskokkos_2025} interface, again rank among the fastest potentials evaluated.

\section{Scaling}\label{sec:scaling}
\begin{figure*}[htbp]
    \centering
    \includegraphics[width=\linewidth]{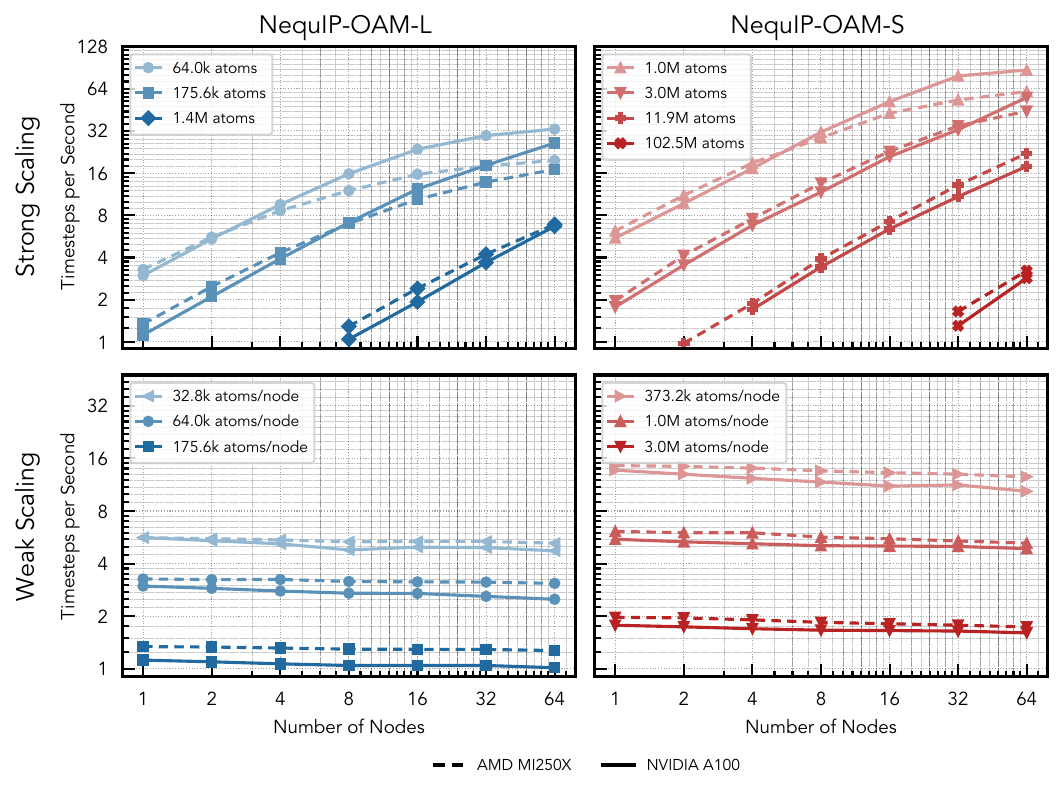}
    \caption{\textbf{Multi-GPU Scaling with NequIP.} Strong (top) and weak (bottom) scaling performance of the large and small NequIP models (NequIP-OAM-L and NequIP-OAM-S) in \texttt{LAMMPS}\cite{thompson_lammps_2022} using the ML-IAP interface\cite{johansson_lammpskokkos_2025} with OpenEquivariance\cite{bharadwaj_efficient_2025} kernels and \texttt{torch.compile}. 
    MD simulations were performed using diamond-structured silicon systems ranging from 32.8k to 102.5M atoms, with NVIDIA A100 (80 GB) and AMD MI250X GPUs on the Perlmutter and Frontier clusters respectively. Each Perlmutter node contains 4 GPUs while each Frontier node contains 8 MI250X GCDs.
    }
    \label{fig:nequip_oam_scaling}
\end{figure*}

To evaluate the utility of our NequIP foundation potentials for large-scale simulations, we evaluated the parallel performance of the large and small (NequIP-OAM-L and NequIP-OAM-S) models in multi-GPU \texttt{LAMMPS} MD simulations, using 4--256 A100 (80 GB) or MI250X GPUs with diamond-structured silicon systems ranging from 32.8k to 102.5M atoms, as shown in \cref{fig:nequip_oam_scaling}.
While the iterative message-passing architecture of NequIP and derived models previously prevented multi-GPU inference in molecular dynamics, the recently-implemented Kokkos-accelerated ML-IAP interface\cite{nguyen-cong_billion_2021,thompson_spectral_2015,johansson_lammpskokkos_2025} with ghost atom exchange now allows efficient parallelisation over many GPUs, along with OpenEquivariance kernels\cite{bharadwaj_efficient_2025} for faster inference.
Under strong scaling (where number of atoms is kept fixed), throughput increases substantially with node count for all system sizes and models, with larger systems exhibiting better parallel efficiency and continuing to scale well up to 64 nodes (256 GPUs). 
Under weak scaling (where atoms/node is kept fixed), throughput remains nearly constant with increasing node counts, indicating ideal scalability and relatively small growth of parallel overhead across the measured range, despite the communication at each layer in the message-passing model.
Scaling performance is closer to ideal for larger system sizes as expected, while A100 (80 GB) GPUs demonstrate greater parallel efficiency over MI250X at large node counts, particularly for smaller systems.
Both models support a large number of atoms before exhausting GPU memory. 
Over our reported datapoints, we show a maximum scale of 22.0k atoms per MI250X GCD on Frontier and 43.9k atoms per A100 (80 GB) GPU on Perlmutter, using NequIP-OAM-L. With NequIP-OAM-S, we can perform a 100M atom simulation with as few as 32 Frontier (AMD) or Perlmutter (NVIDIA) GPU nodes.
The multi-GPU inference scaling performance of Allegro models has already been well demonstrated.\cite{tan_highperformance_2026,kozinsky_scaling_2023,musaelian_learning_2023}

\section{Conclusion}
We have presented a family of equivariant foundation potentials in the NequIP\cite{batzner_e3equivariant_2022} and Allegro\cite{musaelian_learning_2023} architectures, spanning a range of model sizes and trained on inorganic materials datasets of increasing scale and diversity (DIRECT,\cite{qi_robust_2024} MPtrj,\cite{deng_chgnet_2023} MPA,\cite{schmidt_improving_2024} OAM\cite{barroso-luque_open_2024}).
These models achieve competitive accuracies across a range of established community benchmarks -- materials discovery,\cite{riebesell_framework_2025} thermal conductivity,\cite{pota_thermal_2026} and near-equilibrium properties\cite{kaplan_foundational_2025} -- while dominating the accuracy--speed Pareto front.
Dimensionality reduction analysis of chemical type embeddings, shown in \cref{SI:Embedding_Analysis}, demonstrates that fundamental chemical relationships are learned without explicit programming, with progressively clearer chemical structure at larger model sizes.
Large-scale foundation potentials are already supporting diverse, complex machine-learning workflows.
This library of fast and accurate pre-trained models in the widely-used NequIP\cite{batzner_e3equivariant_2022} and Allegro\cite{musaelian_learning_2023} architectures will provide useful starting points for model fine-tuning, greatly aiding data efficiency and accuracy for researchers in the atomistic modelling community. \\

These results were made possible by recent architectural and infrastructural developments in the NequIP framework,\cite{tan_highperformance_2026} including train-time graph compilation, accelerated tensor-product kernels,\cite{bharadwaj_efficient_2025,noauthor_accelerate_2024} judicious mixed-precision training and distributed data-parallel strategies.
Together, these advances yield 5--10x cumulative train-time speed-ups and reduce the cost of training a high-accuracy foundation potential to a few hundred GPU-hours, now within the reach of individual research groups.
Combined with the recently-implemented Kokkos-accelerated ML-IAP interface\cite{johansson_lammpskokkos_2025} in \texttt{LAMMPS},\cite{thompson_lammps_2022} which enables efficient multi-GPU inference for message-passing models, these advances open the door to routine simulation of $\sim$100M-atom systems on relatively modest node counts.
Moreover, the updated NequIP framework prioritises extensibility, allowing for methodological extensions and novel architectures\cite{kim_universal_2025,maruf_learning_2025,falletta_unified_2025,nascimento_mixture_2026,gomes_machinelearned_} or training strategies to readily build on this core framework (to avoid redundant efforts) and fully leverage this accelerated, modular and natively-parallel infrastructure.
\\

Our analysis of error distributions indicates that further progress in MLIPs for materials discovery will depend less on raw model or dataset scaling and more on \emph{improved} data -- broader chemical and structural diversity beyond ionic substitution, consistent (or principled multi-fidelity) treatment of $d$/$f$-electron systems -- and \emph{improved} architectures -- including spin states and accounting for the greater radial and angular resolution demands of correlated systems.
Architecturally, models capable of smoothly varying their radial and angular resolution with the local chemical environment could help reconcile the disparate complexities of various elements/compounds without uniformly inflating model capacity.

\section{Methods}\label{sec:methods}
The full training hyperparameters and package files for all models presented in this work are available through \url{https://nequip.net} and at \url{doi.org/10.5281/zenodo.16980199}, which can be readily downloaded and employed for fine-tuning applications with NequIP and Allegro.
The key hyperparameters varied for different model sizes in this work are shown in \cref{tab:model_hyps}.
In all cases, the parity of feature vector irreducible representations (irreps) was restricted to that of the spherical harmonics (i.e. even for $l = 0, 2...$, odd for $l = 1, 3...$; corresponding to \texttt{parity: false}), which was found to boost speed ($\sim$1.5-2x) with minor effect on accuracies ($<$\SI{3}{\%}).
We note that this choice still corresponds to an $SO(3)$-equivariant network architecture.
Models were trained using PyTorch version 2.7.1, NequIP version 0.16 from \href{https://github.com/mir-group/nequip}{github.com/mir-group/nequip}, Allegro version 0.7 from \href{https://github.com/mir-group/allegro}{github.com/mir-group/allegro}, \texttt{e3nn} version 0.5.6 from \href{https://github.com/e3nn/e3nn}{github.com/e3nn/e3nn}, \texttt{OpenEquivariance} version 0.4 from \href{https://github.com/PASSIONLab/OpenEquivariance}{github.com/PASSIONLab/OpenEquivariance} and \texttt{cuEquivariance} version 0.6 from \href{https://github.com/NVIDIA/cuEquivariance}{github.com/NVIDIA/cuEquivariance} -- though we note that the packaged models (used for fine-tuning) are fully compatible with the latest versions of each. \\

As mentioned above, the use of reduced floating-point precision (such as \texttt{TensorFloat32} format) at \emph{inference} time was found to introduce significant noise in predictions, most notably with forces. 
For the materials discovery benchmark, this increased the number of ionic steps required for geometry relaxations, while the final energies were mostly unaffected.
For the thermal conductivity benchmark, the noise in forces gave dramatically increased predictive errors, as discussed in Refs. \citenum{rhodes_orbv3_2025,fu_learning_2025}.
Increasing the atomic displacement distance in the phonon calculations partially reduced the impact of lower precision on benchmark accuracies.\cite{rhodes_orbv3_2025,fu_learning_2025}
Increasing from full single-precision (\texttt{float32}) to double-precision (\texttt{float64}) at inference did not further improve predictive accuracies, as expected.\cite{kavanagh_identifying_2025,rhodes_orbv3_2025} \\

To support efficient training on large datasets up to terabytes in file sizes, while using distributed data parallel (multiple GPUs), the Lightning Memory Mapped Database (LMDB) format was employed,\cite{_lmdb_} as implemented in the NequIP architecture.\cite{tan_highperformance_2026} 
We note that the total number of parameters in our NequIP models is quite sensitive to the number of node (atom) embedding features (\verb|type_embed_num_features|), but with relatively minor effect on speed and accuracy -- once again highlighting that model parameter count can only be taken as a very rough estimate of the order of magnitude of training and inference speed.
Adaptive loss weighting was trialled, where loss coefficients are dynamically updated during training based on validation error updates, but did not yield a significant improvement in test accuracies for these large datasets. 
This included an extension of the approach proposed by Ocampo et al.\cite{ocampo_adaptive_2024} -- which tends to a 1:1:1 energy:force:stress loss coefficient ratio -- to weight the updates by the initial loss coefficients, but still did not give meaningful improvements.
This approach may be more beneficial when a suitable choice of loss coefficients is not known \textit{a priori}, however.\\

In addition to the results shown in \cref{fig:train_times}, training speeds evaluated using the DIRECT\cite{qi_robust_2024} dataset are also provided in \cref{sifig:DIRECT_train_times}, showing the same trends.
For all training speed measurements, a per-GPU batch size of 40 frames was used for NequIP (giving a total batch size of 160), with the `large' model configuration (\cref{tab:model_hyps}).
For Allegro, a reduced batch size of 6 frames per-GPU (total batch size of 24), with $l_{\textrm{max}} = 2$, \SI{5}{\AA} radial cutoff and 4 layers was used to prevent memory overload for the baseline models.
We note that the exact speedup factors from graph compilation, accelerated tensor product kernels and mixed precision will depend on the set of model hyperparameters, which influence the fraction of time spent on different numerical operations such as matrix multiplications, tensor products and so on.
Train-time speed results for NequIP with automatic mixed-precision were not recorded as this precision setting is incompatible with \texttt{OpenEquivariance}\cite{bharadwaj_efficient_2025} in PyTorch $\leq2.7.1$, but is expected to be supported in the latest PyTorch releases.\\

Recent work\cite{kuryla_how_2025} has revealed non-zero net forces in subsets of large DFT datasets, including OMat24\cite{barroso-luque_open_2024}, a signature of imperfect convergence in electronic densities and forces.
We trialled training on only frames with net force (drift) less than 5 or \SI{10}{\%} of the summed force magnitude (as a measure of potential relative error), which corresponded to the removal of a few percent of the \textit{ab initio} MD sub-dataset of OMat24,\cite{barroso-luque_open_2024} but was found to have a negligible effect on model accuracies -- in line with other observations.\cite{barroso-luque_open_2024}
Notably, the total energies are found to be relatively unaffected by under-convergence in these configurations.\cite{sahoo_insights_2026}
We also trialled training on \emph{only} the \textit{ab initio} MD sub-dataset of OMat24,\cite{barroso-luque_open_2024} as in Ref.~\citenum{rhodes_orbv3_2025} where it was found to improve homo-nuclear diatomic energy curves, but again it was found to have mostly negligible effect (beyond a slight reduction in materials discovery accuracy), while NequIP models trained on the full OMat24\cite{barroso-luque_open_2024} dataset show good homo-nuclear energy accuracies.\cite{chiang_mlip_2025}\\

For the MatCalc\cite{kaplan_foundational_2025} geometry relaxation benchmarks, the initial structure perturbation was re-ran with the same random seed (=8) for all models, to ensure a like-for-like comparison. 
Convex hull energy mean absolute errors ($E_{\textrm{WBM, MAE}}$) are taken over the set of unique structural prototypes from the WBM test set ($\sim$215k of the $\sim$250k compounds).\cite{riebesell_framework_2025}
To decompose $E_{\textrm{WBM, MAE}}$ to per-element values in \cref{fig:per-element_errors}, the mean value of absolute errors weighted by atomic fraction was used:
\begin{equation}
    \mathrm{MAE}_X = \frac{\sum_i f_{X,i} \, \left|E_{\textrm{WBM, i}}\right|}{\sum_i f_{X,i}},
\end{equation}
where $f_{X,i} = N_{X,i}/N_{\mathrm{atoms},i}$ is the atomic fraction of element $X$ in structure $i$, with the sum running over the full WBM\cite{wang_predicting_2021} test set.
Xe is omitted from the element-wise error distribution in \cref{fig:per-element_errors} as it only occurs once (with \ce{XeF2}) in the WBM\cite{wang_predicting_2021} test set of $\sim$250k compounds, and thus is not statistically significant.\\

Model training was primarily performed using NVIDIA A100/H100/H200 GPUs at the FASRC Cannon cluster at Harvard University and the NERSC Perlmutter system.
A total batch size of 640 frames was used, distributed over 4--16 GPUs.
Total model training costs varied from $\sim$100 to 750 (H200) GPU hours.
The AdamW optimiser was used in all cases, with a \texttt{ReduceLROnPlateau} learning-rate schedule. 
The energy:force:stress (E:F:S) loss function coefficients were initialised to 1:5:0.1, with training continued until force validation errors plateaued, before a second training stage with 1:1:0.1 E:F:S weighting, (initial) learning rate reduced by a factor of 10 and no mixed-precision.
Per-atom energy loss was used, rather than total energy.
For the NequIP/Allegro-OAM-\texttt{\{model size\}} models, an initial learning rate of 0.005 and weight decay of 1e-8 was used with training on OMat24\cite{barroso-luque_open_2024} only, before running the second (fine-tuning) training stage on the MPA dataset.
An initial learning rate and weight decay of 0.005 and 1e-8 was used when training on MPA, while 0.02 and 1e-3 was used with MPtrj/DIRECT datasets.
Clipping of gradient norms to 1 was used when training on OAM and MPA, while 0.01 was used with MPtrj and DIRECT.
Huber loss functions were used, with delta values of \SI{0.01}{eV/atom} for energy, \SI{0.01}{eV/\AA} for force and \SI{0.1}{eV/\AA^3} for stress.
A stratified Huber loss function was also tested for forces, using successively smaller Huber delta values for extreme force magnitudes ($>$\SI{100}{eV/\AA}).
This was previously found to aid training stability with the MPtrj\cite{deng_chgnet_2023} dataset,\cite{batatia_foundation_2024} but appeared to have negligible effect with our setup.
Isolated atom energies were used for the reference atomic energy shifts in model training, taken from \url{https://github.com/esoteric-ephemera/isolated_atomic_energies} and \url{https://github.com/facebookresearch/fairchem/blob/main/configs/uma/training_release/element_refs/iso_atom_elem_refs.yaml}, while the square-root of per-element root-mean-square force magnitudes on MPtrj\cite{deng_chgnet_2023} was used for initial model energy scales.\\

The FIRE\cite{bitzek_structural_2006} optimiser was used for geometry relaxations when evaluating the \texttt{matbench-discovery}\cite{riebesell_framework_2025} materials discovery accuracy, along with a maximum ionic force convergence threshold of \SI{0.015}{eV/\AA} and a maximum of 200 relaxation steps.
A displacement distance of \SI{0.035}{\AA} was used for the finite-difference phonon calculations when computing the $\kappa_{\textrm{SRME}}$\cite{pota_thermal_2026} accuracies.\\

The LAMMPS scaling tests performed on Frontier and Perlmutter utilised the 30 March 2026 feature release of LAMMPS. The initial silicon structure was obtained from the Materials Project \cite{jain_materials_2013}. On NERSC Perlmutter, all simulations were run on NVIDIA A100 GPUs with 80 GB of HBM2e memory. The software environment consisted of Python 3.11.0, PyTorch 2.9.0+cu129, NequIP 0.17.1, NequIP-Allegro 0.8.2, e3nn 0.6.0, and OpenEquivariance 0.6.6, together with CUDA Toolkit 12.9. The Cray Programming Environment module \texttt{craype-accel-nvidia80} and \texttt{cray-mpich/9.0.1} were used to enable GPU acceleration and MPI communication. On OLCF Frontier, simulations used the AMD MI250X GPUs which have 128GB of HBM2e memory split across two GCDs, in a software environment with Python 3.11.15, PyTorch 2.11.0+ROCm7.2, NequIP 0.16.1, NequIP-Allegro 0.8.0, e3nn 0.5.9, and OpenEquivariance 0.6.6. The key system modules \texttt{rocm/7.2.0}, \texttt{PrgEng-amd/8.6.0}, and \texttt{cray-mpich/9.1.0} were activated. \\

Both \texttt{ASE}\cite{larsen_atomic_2017} and \texttt{LAMMPS}\cite{thompson_lammps_2022} molecular dynamics inference speed tests were performed using the same diamond silicon structure as the \texttt{LAMMPS}\cite{thompson_lammps_2022} scaling tests, though at smaller supercell sizes.
These single-GPU speed comparisons were performed on one NVIDIA A100-SXM4 (80\,GB) GPU, using the publicly-documented setup and available accelerators for each model.
When evaluating the \texttt{ASE}\cite{larsen_atomic_2017} and \texttt{LAMMPS}\cite{thompson_lammps_2022} MD runtimes, simulations were run in the $NVE$ ensemble with a \SI{1}{fs} timestep, for 1{,}000 unrecorded warm-up steps followed by 5{,}000 timed steps (200 warm-up and 500 timed steps for the substantially more expensive PET-OAM-XL\cite{mazitov_petmad_2025,bigi_pushing_2026} model, with \texttt{LAMMPS}\cite{thompson_lammps_2022}).
Reduced-precision (\texttt{TensorFloat32}) arithmetic was disabled in all cases, consistent with the inference settings used elsewhere in this work, except for a subset of the PET-OAM-XL \texttt{LAMMPS}\cite{thompson_lammps_2022} benchmarks (\cref{sifig:lammps_speeds}).
While inference speeds were recorded for both conserving and non-conserving variants of PET-OAM-XL,\cite{mazitov_petmad_2025,bigi_pushing_2026} only the conserving model speeds were used for comparisons with the other (conserving) foundation MLIPs in \cref{fig:pareto}.
For NequIP, Allegro and SevenNet, which support multiple \texttt{LAMMPS}\cite{thompson_lammps_2022} interfaces (ML-IAP, Pairstyle, e3gnn), the fastest inference speed recorded for a given model and supercell size with any interface was used for \cref{fig:pareto}, while the full set of inference speeds is provided in \cref{sifig:lammps_speeds}.\\

For the \texttt{ASE}\cite{larsen_atomic_2017} evaluations, ahead-of-time (AOTInductor) compilation via \texttt{nequip-compile} with PyTorch 2.10.0 and accelerated tensor-product kernels were used for NequIP (\texttt{OpenEquivariance}\cite{bharadwaj_efficient_2025} kernels) and Allegro (\texttt{cuEquivariance}\cite{noauthor_accelerate_2024} kernels).
\texttt{cuEquivariance}\cite{noauthor_accelerate_2024} tensor product kernels were also used with MACE-MPA-0\cite{batatia_foundation_2024}, SevenNet-Omni\cite{kim_optimizing_2025} was accelerated by FlashTP kernels, and Nequix MP PFT\cite{koker_pft_2026} was executed using a JAX backend configured with \texttt{OpenEquivariance} kernels.
\texttt{model.compile()} was used to accelerate ORB v3\cite{rhodes_orbv3_2025}. 
Both eSEN-30M-OAM\cite{fu_learning_2025} and EquiformerV3\cite{liao_equiformerv3_2026} were evaluated via the FairChem/OCP calculator path, PET-OAM-XL\cite{mazitov_petmad_2025,bigi_pushing_2026} via the \texttt{UPETCalculator}, and GRACE\cite{bochkarev_graph_2024,lysogorskiy_graph_2026} models using the \texttt{TensorPotential} calculator with TensorFlow GPU support.\\

To enable direct comparison while accommodating the incompatible software dependencies of each model family, four separate \texttt{LAMMPS} builds with matched Python environments were prepared; with each \texttt{LAMMPS} executable compiled using GCC 12.2 and CUDA 12.9.1, and all runs using a single MPI rank.
The NequIP and Allegro models were run from a single \texttt{LAMMPS} build (release \texttt{patch\_30Mar2026}) with the Kokkos\cite{johansson_lammpskokkos_2025} and ML-IAP packages, linked against PyTorch 2.9.1 (CUDA 12.6) with NequIP 0.17.1 and Allegro 0.8.2. Each model was benchmarked through two interfaces: (i) the native pair styles, exported as ahead-of-time (AOTInductor) compiled artifacts via \texttt{nequip-compile}, and (ii) the Kokkos-accelerated ML-IAP interface,\cite{johansson_lammpskokkos_2025} for which the accelerated tensor-product kernels are integrated --- \texttt{OpenEquivariance}\cite{bharadwaj_efficient_2025} for NequIP and \texttt{cuEquivariance}\cite{noauthor_accelerate_2024} for Allegro. The NequIP pair style has no Kokkos implementation and was therefore run on a single GPU without Kokkos (its only Kokkos-accelerated path being ML-IAP), while the Allegro pair style and both ML-IAP paths used the Kokkos GPU backend.

For the comparison models, we used the same \texttt{LAMMPS}/ML-IAP build to run MACE-MPA-0\cite{batatia_foundation_2024} (exported to the ML-IAP unified format with \texttt{mace\_create\_lammps\_model}), and dedicated \texttt{LAMMPS} builds for GRACE\cite{bochkarev_graph_2024,lysogorskiy_graph_2026} and SevenNet\cite{kim_optimizing_2025} reflecting their distinct interfaces. GRACE-1L and GRACE-2L (small, medium and large variants) were evaluated through the TensorFlow-backed \texttt{pair\_style grace} (LAMMPS release 10\,Sep\,2025, ML-PACE package, TensorFlow 2.19), in which GPU execution is managed by the TensorFlow runtime rather than Kokkos. SevenNet-Omni and its larger \texttt{i12} variant were evaluated through two interfaces: the ML-IAP interface (with \texttt{cuEquivariance} kernels), run on the same shared ML-IAP build (release \texttt{patch\_30Mar2026}) as NequIP/Allegro/MACE, into which the unified ML-IAP artifact loads directly; and the TorchScript-based \texttt{pair\_style e3gnn}\cite{park_scalable_2024} (single MPI rank, with FlashTP), run on a dedicated \texttt{LAMMPS} build (release 2\,Aug\,2023) patched with the SevenNet interface and the FlashTP tensor-product kernels.\\

Finally, PET-OAM-XL\cite{mazitov_petmad_2025,bigi_pushing_2026} was evaluated through a dedicated \texttt{LAMMPS} build of the \texttt{metatomic} branch of \texttt{metatensor/lammps}, which provides the \texttt{pair\_style metatomic} interface with a Kokkos GPU backend (\texttt{metatomic/kk}, compiled for the \texttt{sm\_80} architecture). The public \texttt{v1.0.0} checkpoint was exported to a \texttt{metatomic} TorchScript model with \texttt{metatrain}'s \texttt{mtt export}, and the build was linked against PyTorch 2.10.0 (CUDA 12.8) with \texttt{metatomic-torch} 0.1.11, \texttt{metatensor-torch} 0.8.5 and \texttt{metatrain} 2026.2.1. Four configurations were benchmarked: the default energy-conserving model, in which forces are obtained by automatic differentiation of the predicted potential energy, and a non-conservative variant,\cite{bigi_pushing_2026} in which per-atom forces are produced directly by a dedicated output head, each evaluated with reduced-precision (\texttt{TensorFloat32}) tensor-core arithmetic both disabled and enabled. This build was configured without MPI and all runs used a single Kokkos GPU rank. All comparison-model checkpoints were obtained from their respective public releases.\\

Harmonic phonons were computed for the 9958 materials of the MDR phonon database\cite{_mdr_} following the protocol of Ref.~\citenum{loew_universal_2025}. 
Each MDR unit cell was relaxed with the model in \texttt{ASE}\cite{larsen_atomic_2017} using the FIRE\cite{bitzek_structural_2006} optimiser with crystal symmetry constrained and cell volume allowed to vary.
An ionic force convergence threshold of \SI{5e-3}{eV/\angstrom} with 3000 maximum ionic steps was used.
Force constants were then obtained using \texttt{phonopy}\cite{togo_first_2015} with model forces on symmetry-reduced $\pm$ displacements of \SI{0.01}{\angstrom}, using the MDR reference supercell expansions for each material.
No exclusion of materials predicted to be dynamically-unstable is performed, but imaginary modes are excluded from the thermal sums (following the \texttt{phonopy}\cite{togo_first_2015} default behaviour).
For \cref{sitab:mdr-stability}, a structure is classified as dynamically stable when every predicted frequency is real at every commensurate $q$-point, except the three acoustic branches at $\Gamma$ (permitted down to \SI{-50}{K}), matching the procedure of Ref.~\citenum{loew_universal_2025}.
No published source that we could find reports MDR errors for PET-OAM-XL, so its errors are computed here from the per-structure outputs deposited by Ref.~\nocite{han_phononbench_2026}\citenum{han_phononbench_2026}.

For the analysis of $\kappa_{\textrm{SRME}}$ accuracy and its dependence on interaction range in \cref{sifig:IFC_range}, the PBE finite-displacement dataset from the MDR database\cite{_mdr_} for the 103 compounds in the $\kappa_{\textrm{SRME}}$ test set was used.
Reference second- and third-order force constants were produced from this displacement-force data with \textsc{phono3py},\cite{togo_distributions_2015} and thermal conductivities obtained with the same protocol as the $\kappa_{\textrm{SRME}}$ benchmark.
Range-limited force constants were then constructed by two routes; refitting force constants to the same displacement--force data using \textsc{symfc},\cite{seko_projectorbased_2024} restricted to pairs within the given radial cutoff $r_c$, and instead zeroing the exact DFT force constants beyond $r_c$ (and repairing sum rules without returning any weight outside the cutoff) -- both shown in \cref{sifig:IFC_range}.

In addition to model configuration and package files, all data and analysis code generated as part of this work (including inference speed benchmarks with other model architectures) is provided at \url{doi.org/10.5281/zenodo.22836506}.

\section*{Author contributions}
S.R.K.: conceptualisation, methodology, software, investigation, visualisation, writing -- original draft.
C.W.T., M.W., M.L.D., G.d.M.N., U.U., L.Z., F.L. and N.R.: methodology, software, investigation.
A.G. and V.B.: software.
A.J.: software, methodology.
W.C.W. and A.M.: methodology, software, supervision.
B.K.: supervision, project administration, funding acquisition, resources.
All authors contributed to writing -- review and editing.

\section*{Conflicts of Interest}
There are no conflicts of interest to declare.

\section*{Acknowledgements}
The authors acknowledge useful discussions with David Rogers regarding compilation and parallel inference, Zachary Goodwin regarding phonon prediction accuracies, and the cuEquivariance team at NVIDIA for their guidance on the cuEquivariance API.
SRK thanks the Harvard University Center for the Environment (HUCE) for funding a fellowship.

This research used the FASRC Cannon cluster supported by the FAS Division of Science Research Computing Group at Harvard University, as well as resources of the National Energy Research Scientific Computing Center (NERSC), a Department of Energy User Facility using NERSC awards BESERCAP0026720, BESERCAP0032494 and BESERCAP0038433.
Via our membership of the UK's HEC Materials Chemistry Consortium, which is funded by EPSRC (EP/X035859), this work used the ARCHER2 UK National Supercomputing Service (http://www.archer2.ac.uk).
An award of computer time was provided by the INCITE program. This research used resources of the Oak Ridge Leadership Computing Facility, which is a DOE Office of Science User Facility supported under Contract DE-AC05-00OR22725.
This work was supported by the National Science Foundation through the Harvard University Materials Research Science and Engineering Center Grant No. DMR-2011754.
L.Z. was supported by the National Science Foundation Graduate Research Fellowship under Grant No. DGE-2140743.
Sandia National Laboratories is a multi-mission laboratory managed and operated by National Technology \& Engineering Solutions of Sandia, LLC (NTESS), a wholly owned subsidiary of Honeywell International Inc., for the U.S. Department of Energy’s National Nuclear Security Administration (DOE/NNSA) under contract DE-NA0003525. This written work is authored by an employee of NTESS. The employee, not NTESS, owns the right, title and interest in and to the written work and is responsible for its contents. Any subjective views or opinions that might be expressed in the written work do not necessarily represent the views of the U.S. Government. The publisher acknowledges that the U.S. Government retains a non-exclusive, paid-up, irrevocable, world-wide license to publish or reproduce the published form of this written work or allow others to do so, for U.S. Government purposes. The DOE will provide public access to results of federally sponsored research in accordance with the DOE Public Access Plan.

\putbib
\end{bibunit}
\clearpage 
\onecolumngrid 
\setcounter{equation}{0}
\setcounter{figure}{0}
\setcounter{table}{0}
\setcounter{section}{0}
\setcounter{secnumdepth}{3} 
\setcounter{page}{1}
\makeatletter
\renewcommand{\theequation}{S\arabic{equation}}
\renewcommand{\thefigure}{S\arabic{figure}}
\renewcommand{\thetable}{S\arabic{table}}
\renewcommand{\thesection}{S\arabic{section}}
\begin{bibunit}  
\newcommand{\green}[1]{{\leavevmode\color{black}{#1}}}  

\title[Fast and Accurate Equivariant Foundation Models for Atomistic Simulation (SI)]{Supplementary Material: Fast and Accurate Equivariant Foundation Models for Atomistic Simulation}

\section{Additional Train-Time Benchmarks}\label{SI:Additional_Train_Time_Benchmarks}
\begin{figure}[h]
\includegraphics[width=\linewidth]{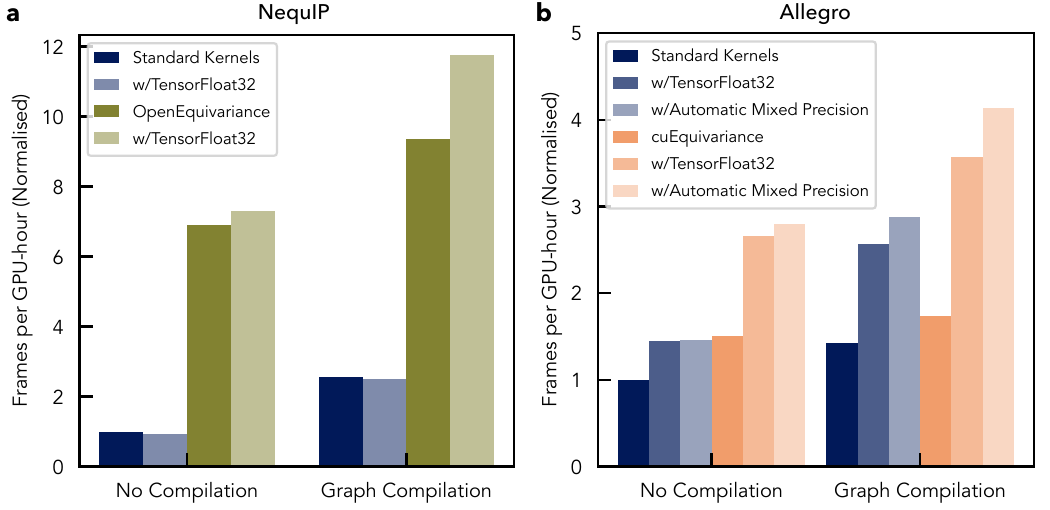}
\caption{\textbf{Accelerated Model Training (with DIRECT dataset).} Training speeds for large NequIP and Allegro models with the DIRECT dataset,\cite{qi_robust_2024} plotted as the number of data frames (atomic structures with energy and forces) trained on per GPU-hour.
Training speeds are measured without and with graph compilation (\texttt{torch.compile}), using standard tensor product kernels or the recently-implemented \texttt{OpenEquivariance}\cite{bharadwaj_efficient_2025} (for NequIP) and \texttt{CuEquivariance}\cite{noauthor_accelerate_2024} (for Allegro) kernels, and with or without mixed floating-point precision strategies (\texttt{TensorFloat32} or Automatic Mixed Precision\cite{_bfloat16_}).
Training times were recorded using a single node of 4 NVIDIA A100 40 GB GPUs on the NERSC Perlmutter system, with further details given in \cref{sec:methods}.
The same batch sizes were used to standardise comparisons, but we note that mixed-precision and accelerated kernels also significantly reduce GPU memory demand, allowing larger batch sizes and thus second-order train-time accelerations.}
\label{sifig:DIRECT_train_times}
\end{figure}

\FloatBarrier 
\section{Additional Learning Curve Analyses}
\begin{figure}[ht]
\centering
\includegraphics[width=0.82\linewidth]{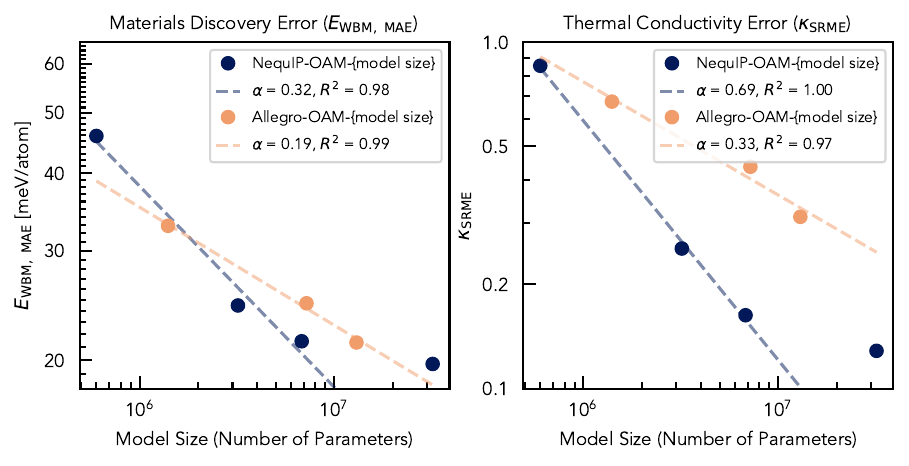}  
\caption{\textbf{Model Size Learning Curve -- Small-to-Large Fit}. 
Test errors for NequIP and Allegro models trained on the OAM dataset (NequIP/Allegro-OAM-\texttt{\{model size\}}) and evaluated on the Matbench Discovery\cite{riebesell_framework_2025} benchmark suite, as a function of model size (\cref{tab:model_hyps,tab:benchmarks}).
This includes the mean absolute error (MAE) in convex hull energy predictions for unseen compounds ($E_{\textrm{WBM, MAE}}$, left) and the dimensionless Symmetric Relative Mean Error in thermal conductivities for 103 binary compounds ($\kappa_{\textrm{SRME}}$, right).
Dashed lines show least-squares power law fits to the data ($L = a\cdot x^{-\alpha}$) -- \textbf{excluding the NequIP-OAM-XL model}, with the fitted scaling exponent $\alpha$ and coefficient of determination $R^2$ given in the figure legends. 
See \cref{tab:benchmarks} for tabulated values.
Model hyperparameters, training datasets and naming syntax are described in \cref{sec:training}.
}
\label{sifig:benchmark_by_model_size_3_data_points}
\end{figure}

\begin{figure}[ht]
\centering
\includegraphics[width=0.82\linewidth]{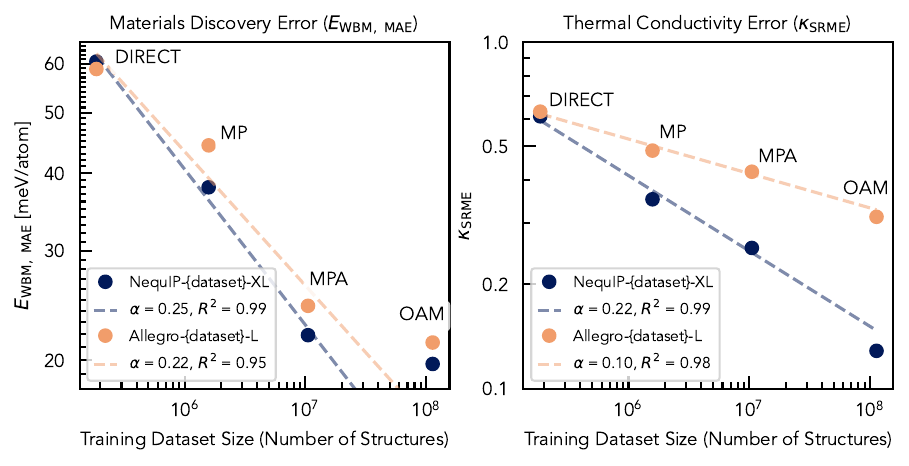}  
\caption{\textbf{Training Data Learning Curve -- DIRECT-to-MPA Fit}. 
Test errors for the largest NequIP and Allegro models trained in this work (NequIP-\texttt{\{dataset\}}-XL and Allegro-\texttt{\{dataset\}}-L) and evaluated on the Matbench Discovery\cite{riebesell_framework_2025} benchmark suite, as a function of training dataset size (DIRECT, MP, MPA, OAM).
This includes the mean absolute error (MAE) in convex hull energy predictions for unseen compounds ($E_{\textrm{WBM, MAE}}$, left) and the dimensionless Symmetric Relative Mean Error in thermal conductivities for 103 binary compounds ($\kappa_{\textrm{SRME}}$, right).
Dashed lines show least-squares power law fits to the data ($L = a\cdot x^{-\alpha}$) -- \textbf{excluding the OAM dataset}, with the fitted scaling exponent $\alpha$ and coefficient of determination $R^2$ given in the figure legends. 
See \cref{tab:benchmarks} for tabulated values and \cref{sec:training} for model hyperparameters, training datasets and naming syntax.}
\label{sifig:benchmark_by_dataset_size_3_data_points}
\end{figure}

\FloatBarrier 
\section{Element-Wise Materials Discovery Errors for Other Community Models}\label{SI:Other_Model_Element_Wise_Errors}
\begin{figure}[htbp]
\centering
\includegraphics[width=0.81\linewidth]{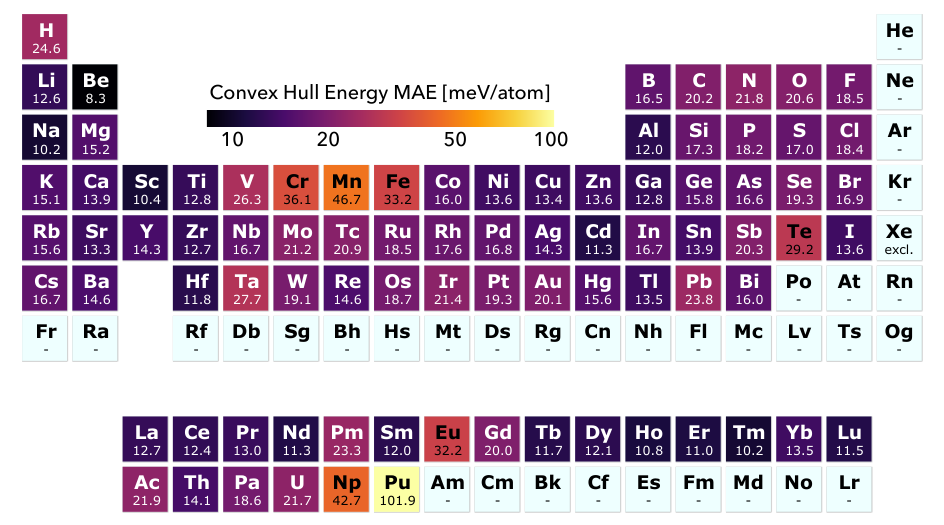}  
\caption{\textbf{Per-Element Errors for Materials Discovery with eSEN-30M-OAM}. 
Distribution of element-wise convex hull energy MAE across the periodic table for the \texttt{matbench-discovery}\cite{riebesell_framework_2025} materials discovery benchmark, using the eSEN-30M-OAM\cite{fu_learning_2025} model.
Errors are shown in meV/atom, using a logarithmic colourbar scale.
The largest element-wise errors are seen for multi-valent first-row transition metals (V, Cr, Mn, Fe), low-prevalence actinides (Np, Pu), along with Eu, Ta, Te, Pb and H -- consistent with the NequIP-OAM models (\cref{fig:per-element_errors}).}
\label{sifig:eSEN_per-element_errors}
\end{figure}

\begin{figure}[htbp]
\centering
\includegraphics[width=0.81\linewidth]{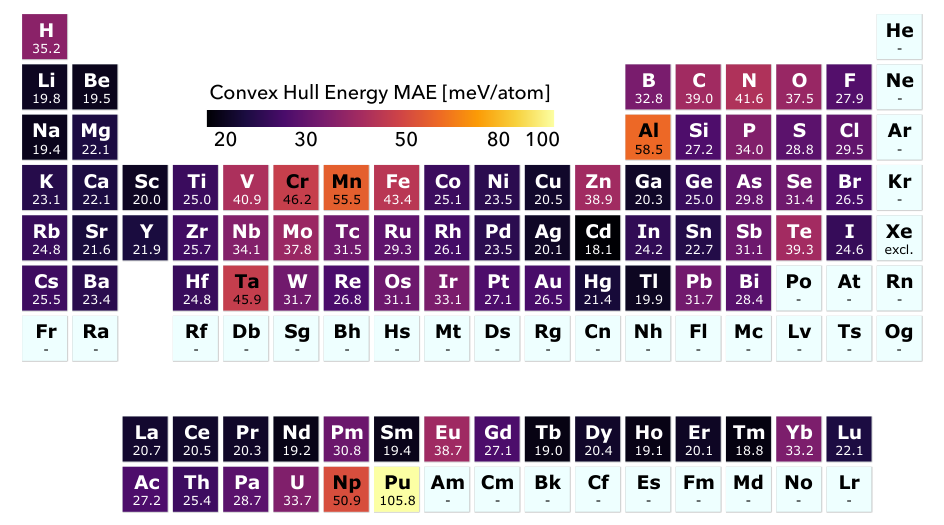}  
\caption{\textbf{Per-Element Errors for Materials Discovery with MACE-MPA-0\cite{batatia_foundation_2024}}.
Distribution of element-wise convex hull energy MAE across the periodic table for the \texttt{matbench-discovery}\cite{riebesell_framework_2025} materials discovery benchmark, using the MACE-MPA-0 model.
Errors are shown in meV/atom, using a logarithmic colourbar scale.
The largest element-wise errors are seen for multi-valent first-row transition metals (V, Cr, Mn, Fe), low-prevalence actinides (Np, Pu), along with Eu, Ta, Te, Pb and H -- consistent with the NequIP-OAM models (\cref{fig:per-element_errors}).
Aluminium has a noticeably higher (relative) error than with NequIP-OAM or eSEN-30M-OAM here.}
\label{sifig:MACE_per-element_errors}
\end{figure}

\FloatBarrier 
\section{Thermal Conductivity and Phonon Benchmark Error Analysis}\label{SI:Thermal_Properties_Error_Analysis}
\begin{figure}[htbp]
\centering
\includegraphics[width=\linewidth]{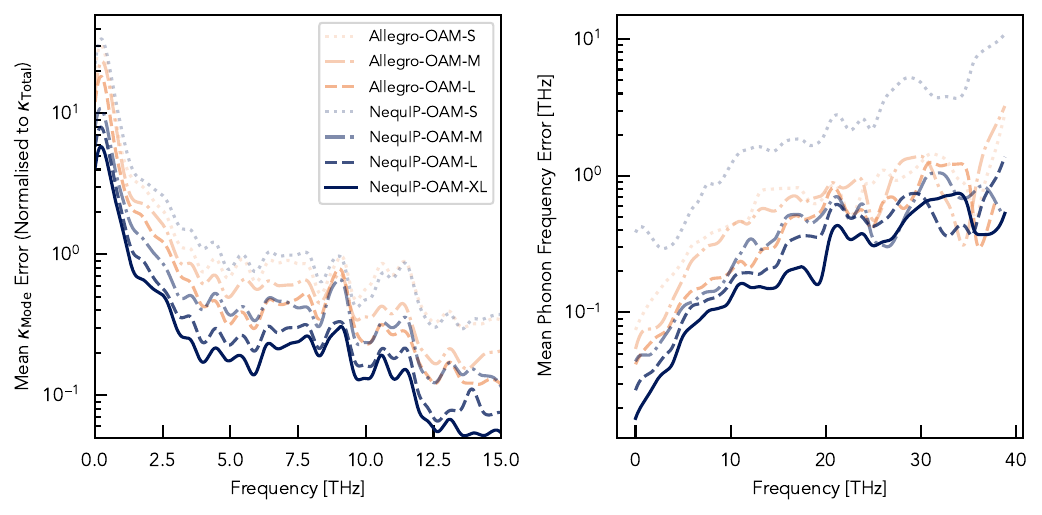}
\caption{\textbf{Decomposed Thermal Conductivity Benchmark Errors}. 
Mean error in \textbf{(left)} mode-decomposed thermal conductivity and \textbf{(right)} phonon frequency for NequIP/Allegro-OAM models over the original $\kappa_{\textrm{SRME}}$ benchmark test set of 103 binary compounds, as a function of phonon frequency.
Thermal conductivities are dominated by low-frequency modes, so the $\kappa_{\textrm{Mode}}$ errors are plotted in the lower energy range of 0 -- \SI{15}{THz}, and are normalised by the total thermal conductivity $\kappa_{\textrm{Total}}$.
Gaussian smoothing of width 0.2 and \SI{0.5}{THz} respectively is applied to the $\kappa_{\textrm{Mode}}$ (left) and phonon frequency (right) error plots.
}
\label{sifig:SRME_Error_Decomposition}
\end{figure}

\begin{figure}[htbp]
\centering
\includegraphics[width=\linewidth]{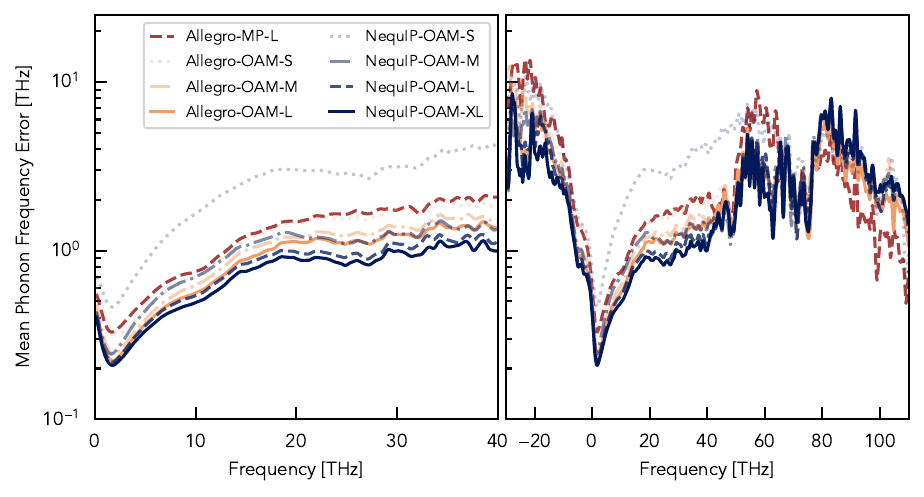}
\caption{\textbf{Mean phonon frequency errors on Materials Project structures.} 
Mean absolute error in phonon frequencies, $|\Delta\omega|$, for NequIP-OAM, Allegro-OAM and Allegro-MP-L models evaluated on all 26,234 Materials Project structures for which reference DFT phonon data is available, as a function of phonon frequency.
The \textbf{left} panel shows the low-frequency range (0 to 40~THz) which encompasses most phonon spectra, while the \textbf{right} panel shows a wider range ($-30$ to 110~THz) for which there are at least 100 datapoints per frequency bin (0.15~THz width).
Errors are averaged over all phonon modes in each frequency bin. 
A displacement distance of \SI{0.001}{\AA} was used for the finite-difference calculations.
\cref{sifig:MP_phonon_errors_smooth} shows the same error distribution but with Gaussian smoothing applied.
}
\label{sifig:MP_phonon_errors_raw}
\end{figure}

\begin{figure}[htbp]
\centering
\includegraphics[width=\linewidth]{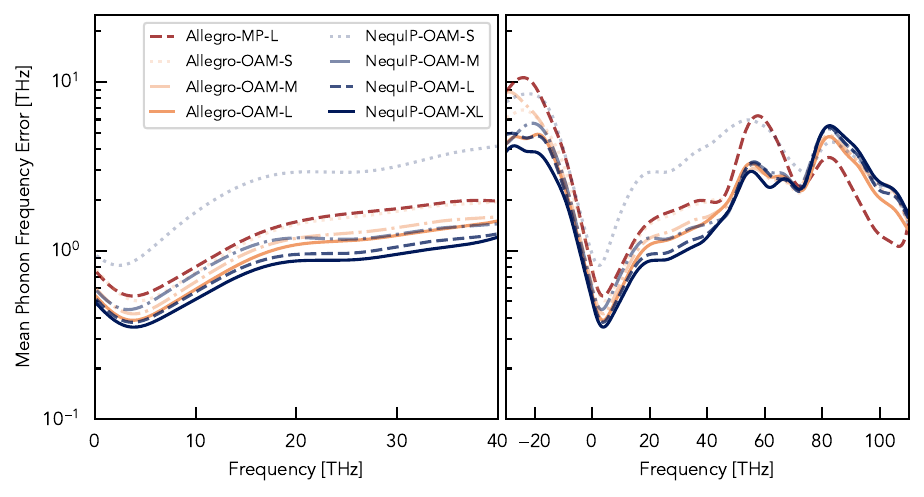}
\caption{\textbf{Mean phonon frequency errors on Materials Project structures (Smoothed).} 
As in \cref{sifig:MP_phonon_errors_raw}, but with Gaussian smoothing $\sigma = 3$~THz applied.
}
\label{sifig:MP_phonon_errors_smooth}
\end{figure}

\begin{figure}[htbp]
\centering
\includegraphics[width=0.9\linewidth]{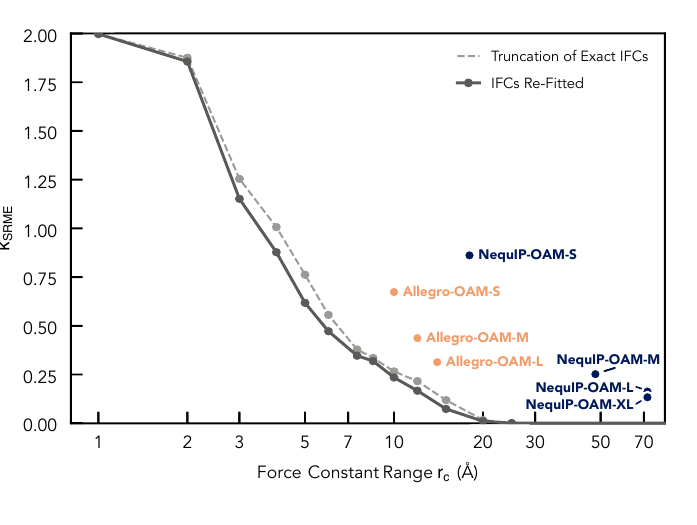}
\caption{\textbf{Thermal conductivity accuracy ($\kappa_{\textrm{SRME}}$) as a function of interaction (interatomic force constant) range.} 
Plot of the minimum achievable ($\kappa_{\textrm{SRME}}$) thermal conductivity benchmark accuracy as a function of the interaction range, where interatomic force constants (IFCs) are re-fitted on reference DFT displacement-force data of atomic pairs up to the radial cutoff $r_c$ (solid grey line), and where the exact DFT IFCs are simply zeroed beyond $r_c$ (dashed grey line).
Alongside are shown the $\kappa_{\textrm{SRME}}$ accuracies achieved by the NequIP and Allegro OAM models, as a function of their interaction ranges (i.e. receptive fields; $2 \times n_{\mathrm{layers}} \times R_{\max}$ and $2 \times R_{\max}$ respectively).}
\label{sifig:IFC_range}
\end{figure}

\begin{figure}[htbp]
\centering
\includegraphics[width=\linewidth]{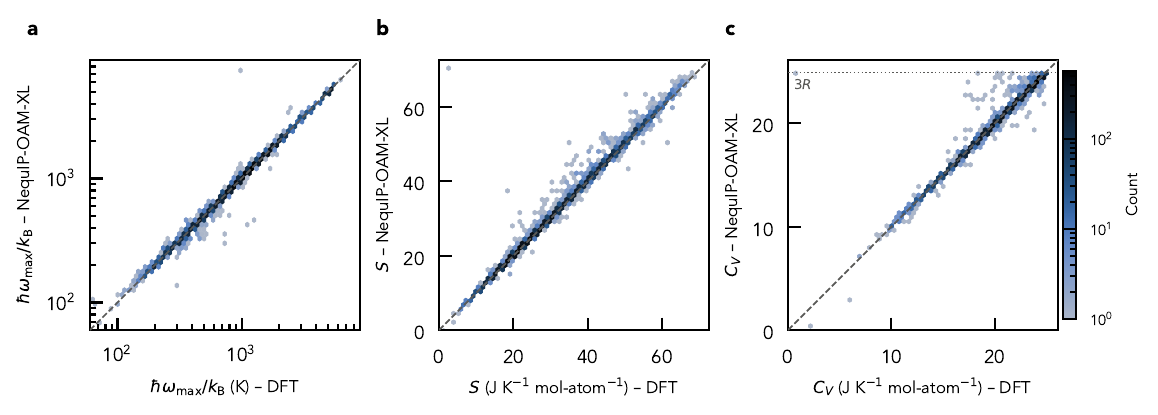}
\caption{\textbf{Harmonic thermal properties benchmark of NequIP-OAM-XL against the MDR DFT phonon database.} Hexagonal-bin parity plots of the highest phonon frequency $\hbar\omega_\mathrm{max}/k_\mathrm{B}$ (\textbf{a}), vibrational entropy $S$ (\textbf{b}) and constant-volume heat capacity $C_V$ (\textbf{c}) over the $\sim$10,000 materials in the MDR database\cite{loew_universal_2025,_mdr_}, computed using the NequIP-OAM-XL model and compared to the PBE DFT reference values.
The dotted line in \textbf{c} is the Dulong--Petit limit $3R = 24.94$~J\,K$^{-1}$\,mol-atom$^{-1}$.
Dashed diagonals are $y=x$. 
Colour gives the number of materials per hexagon on a logarithmic scale (shared colourbar, shown on the right edge).}
\label{sifig:mdr_phonon_parity}
\end{figure}

\begin{figure}[htbp]
\centering
\includegraphics[width=\linewidth]{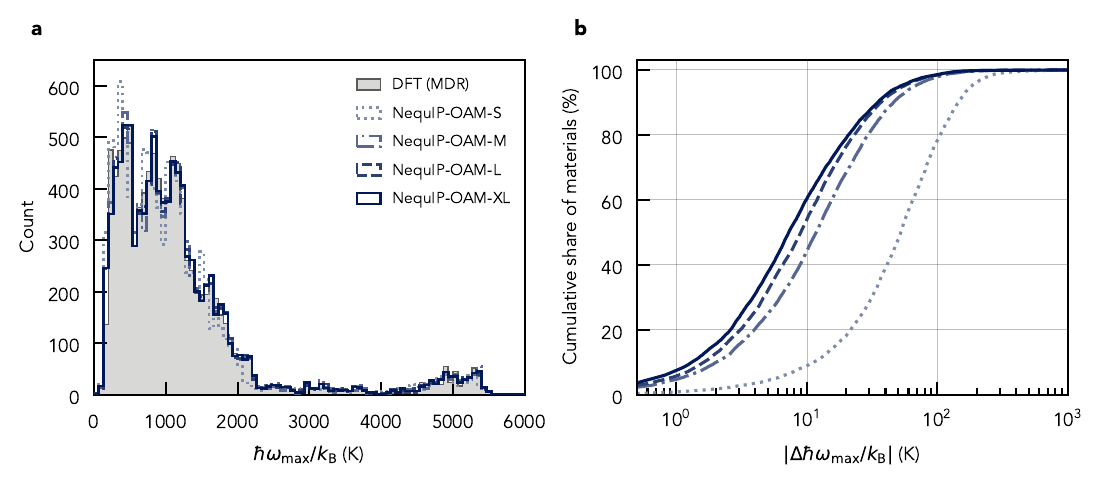}
\caption{\textbf{Highest phonon frequency and its error with NequIP-OAM models across the MDR benchmark.}
(\textbf{a}) Distribution of the maximum phonon frequency $\hbar\omega_\mathrm{max}/k_\mathrm{B}$ over the $\sim$10,000 materials in the MDR dataset, in 67~K bins, for the DFT reference (filled grey histogram) and the four NequIP-OAM models.
\textbf{b}, Cumulative distribution of the absolute error in maximum phonon frequency $|\Delta\hbar\omega_\mathrm{max}/k_\mathrm{B}|$ for each model, on a logarithmic $x$ axis; the value at which each curve crosses 50\% is that model's median error, 55.4, 10.4, 7.2 and \SI{5.6}{K} for S, M, L and XL, against mean absolute errors of 73.3, 16.6, 12.5 and \SI{10.7}{K}.}
\label{sifig:mdr_phonon_distributions}
\end{figure}

\begin{figure}[htbp]
\centering
\includegraphics[width=\linewidth]{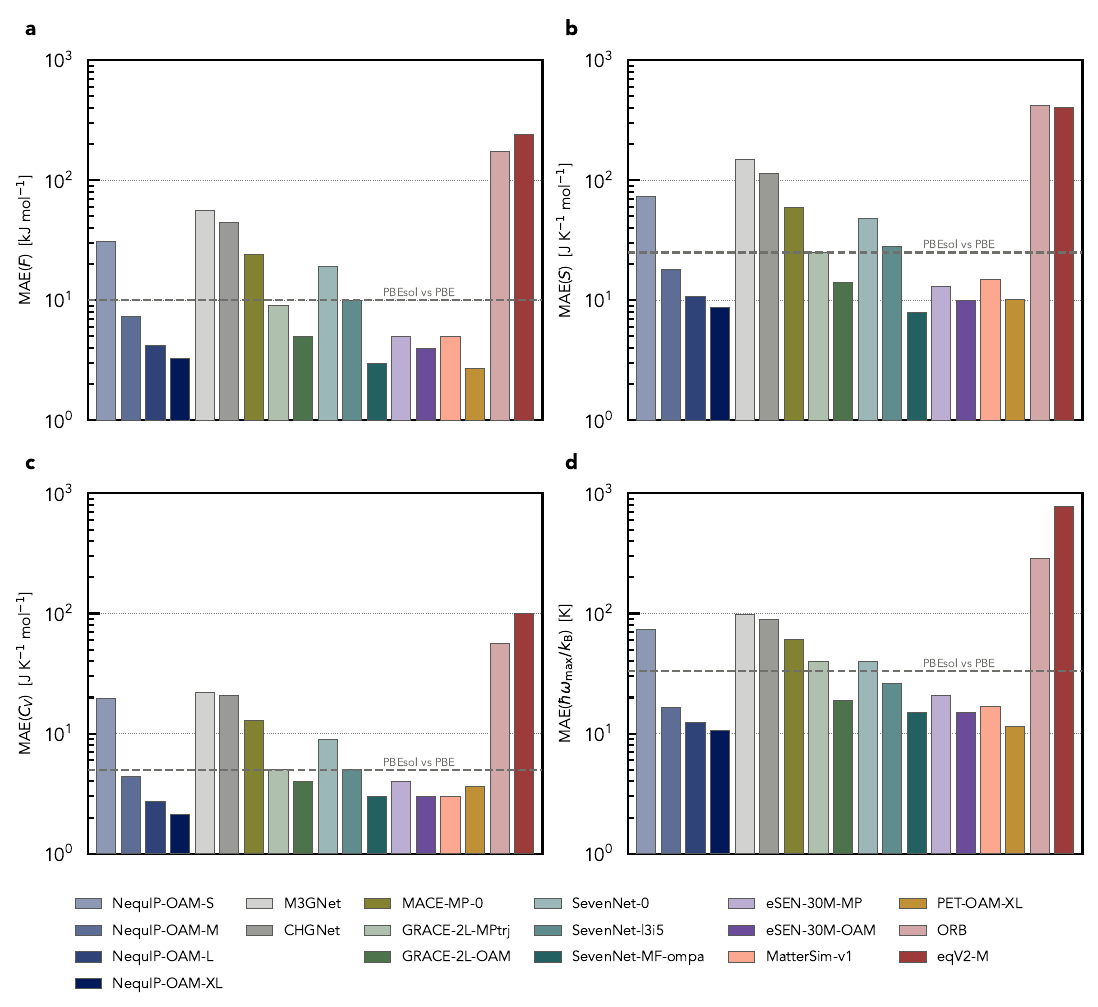}
\caption{\textbf{Mean absolute error in thermal properties on the MDR phonon benchmark for models with published values.} 
\textbf{a}, Vibrational free energy $F$ in kJ\,mol$^{-1}$; \textbf{b}, entropy $S$ in J\,K$^{-1}$\,mol$^{-1}$; \textbf{c}, heat capacity $C_V$ in J\,K$^{-1}$\,mol$^{-1}$, all at \SI{300}{K} and per primitive cell as reported by Refs.~\cite{loew_universal_2025,fu_learning_2025}; \textbf{d}, maximum phonon frequency $\hbar\omega_\mathrm{max}/k_\mathrm{B}$ in K. 
Published values are from Ref.~\citenum{loew_universal_2025} for M3GNet, CHGNet, MACE-MP-0, SevenNet-0, MatterSim-v1, ORB and eqV2-M, and from Ref.~\citenum{fu_learning_2025} for GRACE-2L, SevenNet-l3i5, SevenNet-MF-ompa and the eSEN models. 
PET-OAM-XL values have not been reported but are computed here from the per-structure outputs deposited by PhononBench~\cite{han_phononbench_2026}, scored against the same DFT references.
The dashed line is the PBEsol-vs-PBE functional difference reported in Ref.~\citenum{loew_universal_2025} (where the DFT reference data uses the PBE GGA functional, matching model training data).
ORB and eqV2-M predict forces directly rather than as an energy gradient (i.e. using non-conservative architectures), which is the origin of their poor performance on this thermal properties benchmark (which is sensitive to PES gradients).\cite{loew_universal_2025}}
\label{sifig:mdr_phonon_mae}
\end{figure}

\begin{table*}[htbp]
    \caption{MDR benchmark\cite{_mdr_} results; mean absolute errors in harmonic thermal properties at \SI{300}{K} against the finite-displacement DFT reference, including vibrational free energy $F$, entropy $S$, heat capacity $C_V$ and maximum phonon frequency $\hbar\omega_\mathrm{max}/k_\mathrm{B}$.
    $F$, $S$ and $C_V$ are \emph{per primitive cell}, following Ref.~\citenum{loew_universal_2025} and the \texttt{phonopy}\cite{togo_first_2015} convention. 
    PET-OAM-XL values are computed here from the per-structure outputs deposited by Ref.~\citenum{han_phononbench_2026}. 
    The PBEsol row reports the difference between PBEsol and the PBE reference calculations as reported in Ref.~\citenum{loew_universal_2025}.
    The lowest errors are shown in bold for each property.}
    \centering
    \setlength{\tabcolsep}{7pt}
    \begin{tabular}{llcccc}
        \toprule
        Source & Model & $F_{\textrm{MAE}}$ & $S_{\textrm{MAE}}$ & $C_{V,\ \textrm{MAE}}$ & $\hbar\omega_\mathrm{max}/k_\mathrm{B}$ $_{\textrm{MAE}}$ \\
        & & (kJ\,mol$^{-1}$) & (J\,K$^{-1}$\,mol$^{-1}$) & (J\,K$^{-1}$\,mol$^{-1}$) & (K) \\
        \midrule
        This work & NequIP-OAM-S & 31.1 & 73.1 & 19.8 & 73.3 \\
         & NequIP-OAM-M & 7.29 & 18.2 & 4.41 & 16.6 \\
         & NequIP-OAM-L & 4.21 & 10.8 & 2.74 & 12.5 \\
         & NequIP-OAM-XL & 3.29 & 8.77 & \textbf{2.12} & \textbf{10.7} \\
        \addlinespace
        Ref.~\citenum{loew_universal_2025} & M3GNet & 56 & 150 & 22 & 98 \\
         & CHGNet & 45 & 114 & 21 & 89 \\
         & MACE-MP-0 & 24 & 60 & 13 & 61 \\
         & SevenNet-0 & 19 & 48 & 9 & 40 \\
         & MatterSim-v1 & 5 & 15 & 3 & 17 \\
         & ORB & 175 & 421 & 57 & 291 \\
         & eqV2-M & 241 & 403 & 100 & 780 \\
        \addlinespace
        Ref.~\citenum{fu_learning_2025} & GRACE-2L-MPtrj & 9 & 25 & 5 & 40 \\
         & GRACE-2L-OAM & 5 & 14 & 4 & 19 \\
         & SevenNet-l3i5 & 10 & 28 & 5 & 26 \\
         & eSEN-30M-MP & 5 & 13 & 4 & 21 \\
         & eSEN-30M-OAM & 4 & 10 & 3 & 15 \\
         & SevenNet-MF-ompa & 3 & \textbf{8} & 3 & 15 \\
        \addlinespace
        This work & PET-OAM-XL & \textbf{2.73} & 10.3 & 3.65 & 11.4 \\
        \addlinespace
        Ref.~\citenum{loew_universal_2025} & PBEsol & 10 & 25 & 5 & 33 \\
        \bottomrule
    \end{tabular}
    \label{sitab:mdr-thermal}
\end{table*}

\begin{table*}[htbp]
    \caption{\textbf{Dynamic stability classification across the MDR phonon database}. 
    Accuracy of dynamic stability classification based on phonon spectra computed for the $\sim$10,000 materials in the MDR database, for NequIP-OAM and community models with published values.
    In the MDR DFT reference database, 8239 materials are classified as dynamically stable, while 1719 are classified as dynamically unstable (see \cref{sec:methods}).
    }
    \centering
    \setlength{\tabcolsep}{7pt}
    \begin{tabular}{llcccc}
        \toprule
        Source & Model & True Stable (\%) $\uparrow$ & False Unstable (\%) $\downarrow$ & True Unstable (\%) $\uparrow$ & False Stable (\%) $\downarrow$ \\
        \midrule
        This work & NequIP-OAM-S & 86.7 & 13.3 & 72.4 & 27.6 \\
         & NequIP-OAM-M & 96.1 & 3.9 & 76.4 & 23.6 \\
         & NequIP-OAM-L & 97.1 & 2.9 & 81.2 & 18.8 \\
         & NequIP-OAM-XL & 97.5 & 2.5 & 81.8 & 18.2 \\
        \addlinespace
        Ref.~\citenum{loew_universal_2025} & M3GNet & 87 & 13 & 73 & 27 \\
         & CHGNet & 77 & 23 & 73 & 27 \\
         & MACE-MP-0 & 95 & 5 & 73 & 27 \\
         & SevenNet-0 & 81 & 19 & 80 & 20 \\
         & MatterSim-v1 & 95 & 5 & 75 & 25 \\
         & ORB & 15 & 85 & 92 & 8 \\
         & eqV2-M & 7 & 93 & 94 & 6 \\
        \addlinespace
        Ref.~\citenum{loew_universal_2025} & PBEsol & 97 & 3 & 76 & 24 \\
        \bottomrule
    \end{tabular}
    \label{sitab:mdr-stability}
\end{table*}

\FloatBarrier 
\section{LAMMPS Single-GPU Inference Speed Benchmark}\label{SI:LAMMPS_Speeds}
\begin{figure}[htbp]
\centering
\includegraphics[width=\linewidth]{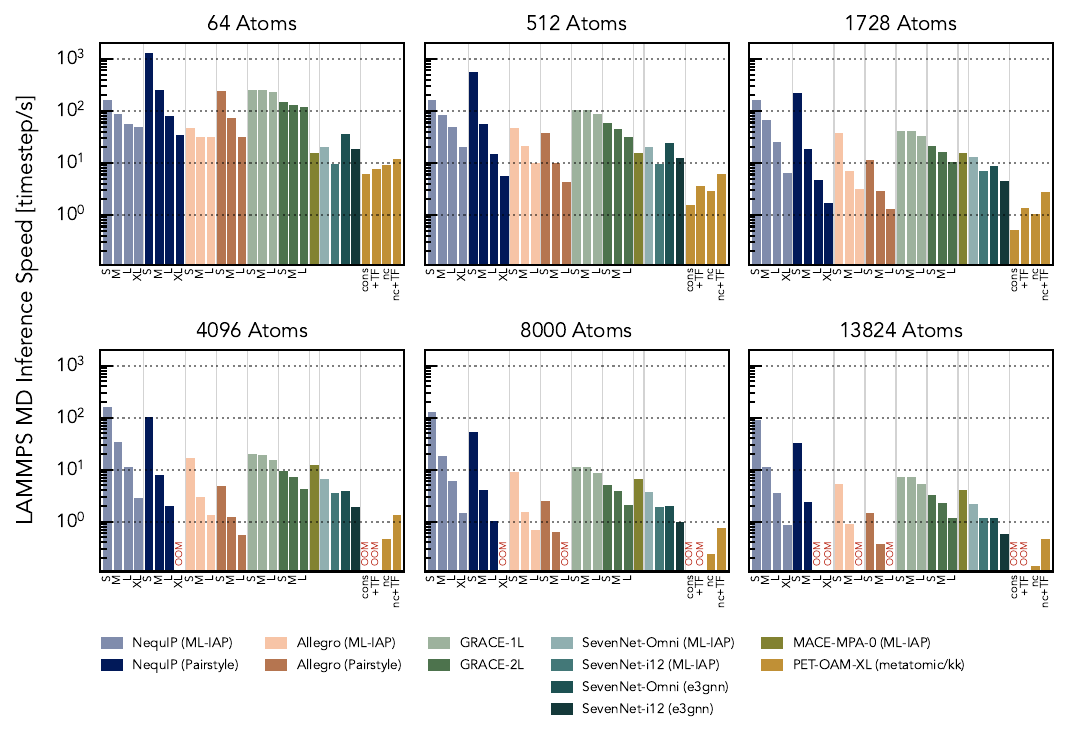}
\caption{\textbf{LAMMPS Single-GPU Inference Speeds}. 
\texttt{LAMMPS}\cite{thompson_lammps_2022} molecular dynamics inference speeds (timesteps per second) for the NequIP-OAM and Allegro-OAM models trained in this work, alongside other community foundation potentials (GRACE-1L/GRACE-2L,\cite{bochkarev_graph_2024,lysogorskiy_graph_2026} MACE-MPA-0,\cite{batatia_foundation_2024} SevenNet-Omni,\cite{kim_optimizing_2025} and PET-OAM-XL\cite{mazitov_petmad_2025,bigi_pushing_2026}), as a function of system size.
Each panel corresponds to a fixed number of atoms (64 to 13{,}824), using diamond silicon (Si) supercells.
NequIP and Allegro models are shown using both the native pairstyle and the Kokkos-accelerated ML-IAP\cite{johansson_lammpskokkos_2025} interfaces.
PET-OAM-XL is run through the Kokkos-accelerated \texttt{metatomic} interface in four variants: the energy-conserving model (``cons'') and its non-conservative, direct-force counterpart (``nc''),\cite{bigi_pushing_2026} each evaluated with and without TF32 tensor-core arithmetic (``+TF'').
All simulations were run on a single NVIDIA A100 (80 GB) GPU (see \cref{sec:methods} of the main text for details).
}
\label{sifig:lammps_speeds}
\end{figure}

\FloatBarrier 
\section{Learned Type Embedding Analysis}\label{SI:Embedding_Analysis}
\begin{figure}[htbp]
\centering
\includegraphics[width=\linewidth]{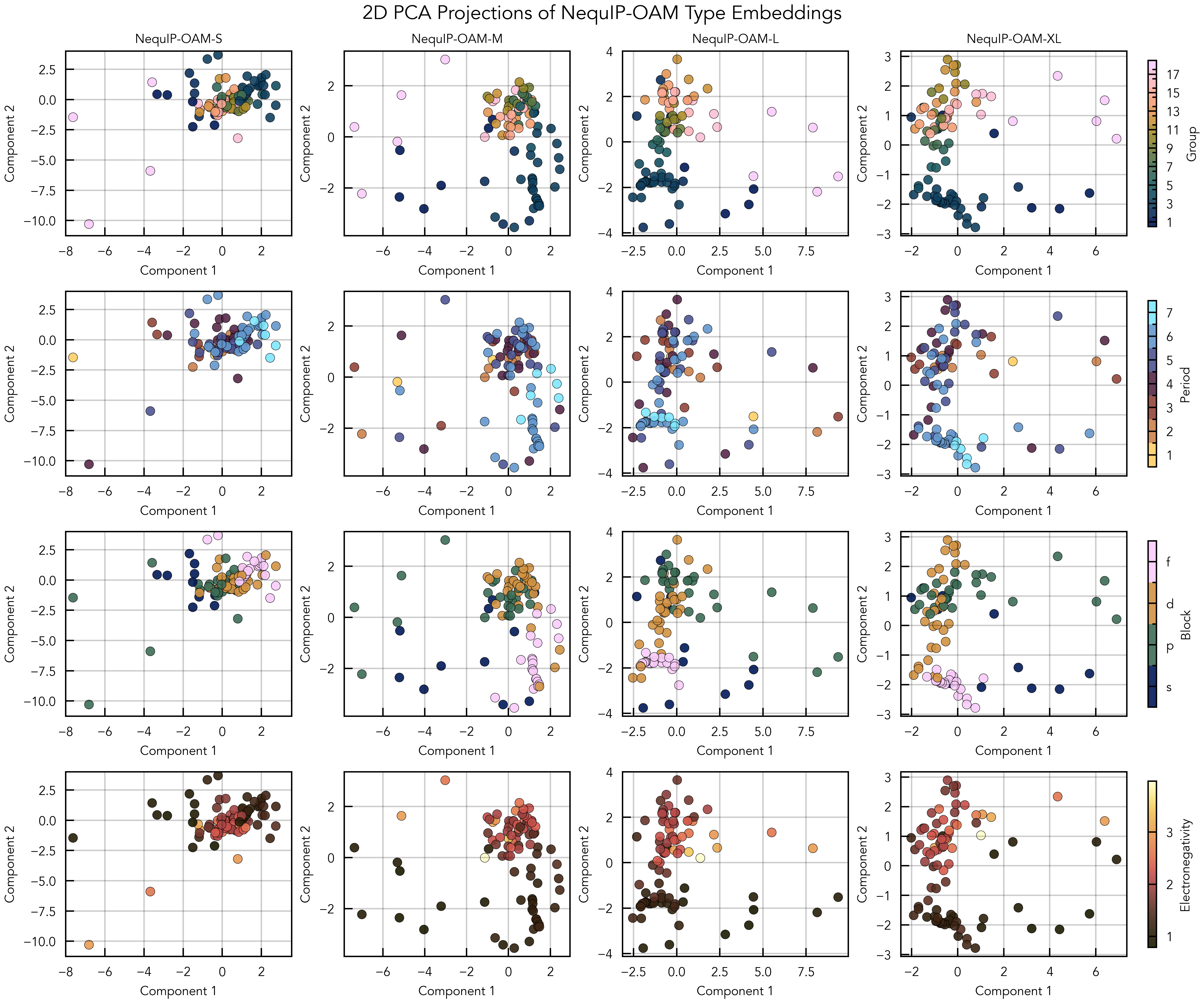}
\caption{\textbf{Principal Component Analysis (PCA) Dimensionality Reduction of Chemical Type Embeddings in NequIP-OAM Models}.
Projections of the chemical type embedding vectors learned by the NequIP-OAM models (S, M, L, XL; columns) along the top 2 principal components.
Scatter points are coloured by periodic group, period (row), periodic-table block, and Pauling electronegativity, in order from the top to bottom row. 
Cumulative explained variance ratios are 0.27, 0.20, 0.16 and 0.18 for the S, M, L and XL models respectively, with the top 2 principal components shown, and 0.48, 0.39, 0.33 and 0.36 for the top 5 principal components.
}
\label{sifig:pca_grid}
\end{figure}

\begin{figure}[htbp]
\centering
\includegraphics[width=\linewidth]{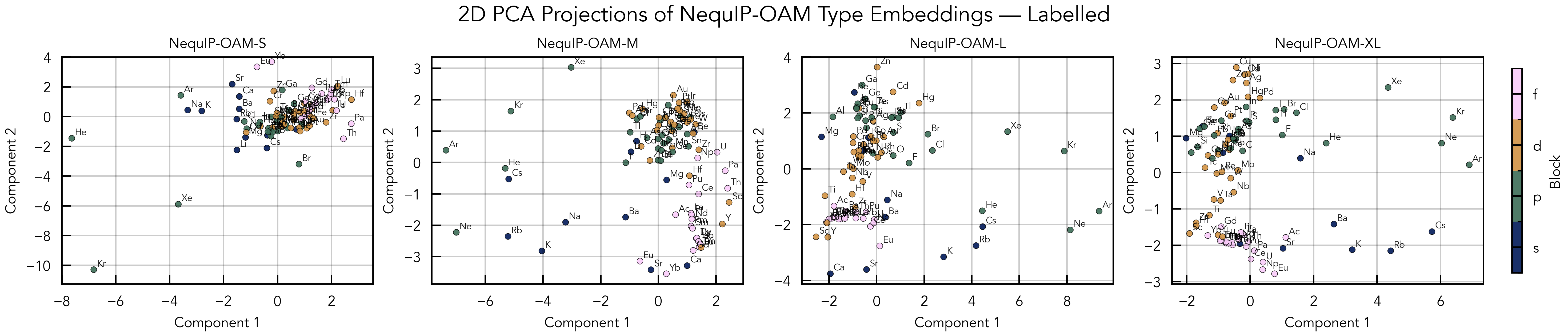}
\caption{\textbf{PCA Dimensionality Reduction of Chemical Type Embeddings in NequIP-OAM Models -- Labelled}.
Projections of the chemical type embedding vectors learned by the NequIP-OAM models (S, M, L, XL; columns) along the top 2 principal components, labelled by the elemental identities.
Scatter points are coloured by periodic-table block.
}
\label{sifig:pca_labeled_all}
\end{figure}

\begin{figure}[htbp]
\centering
\includegraphics[width=\linewidth]{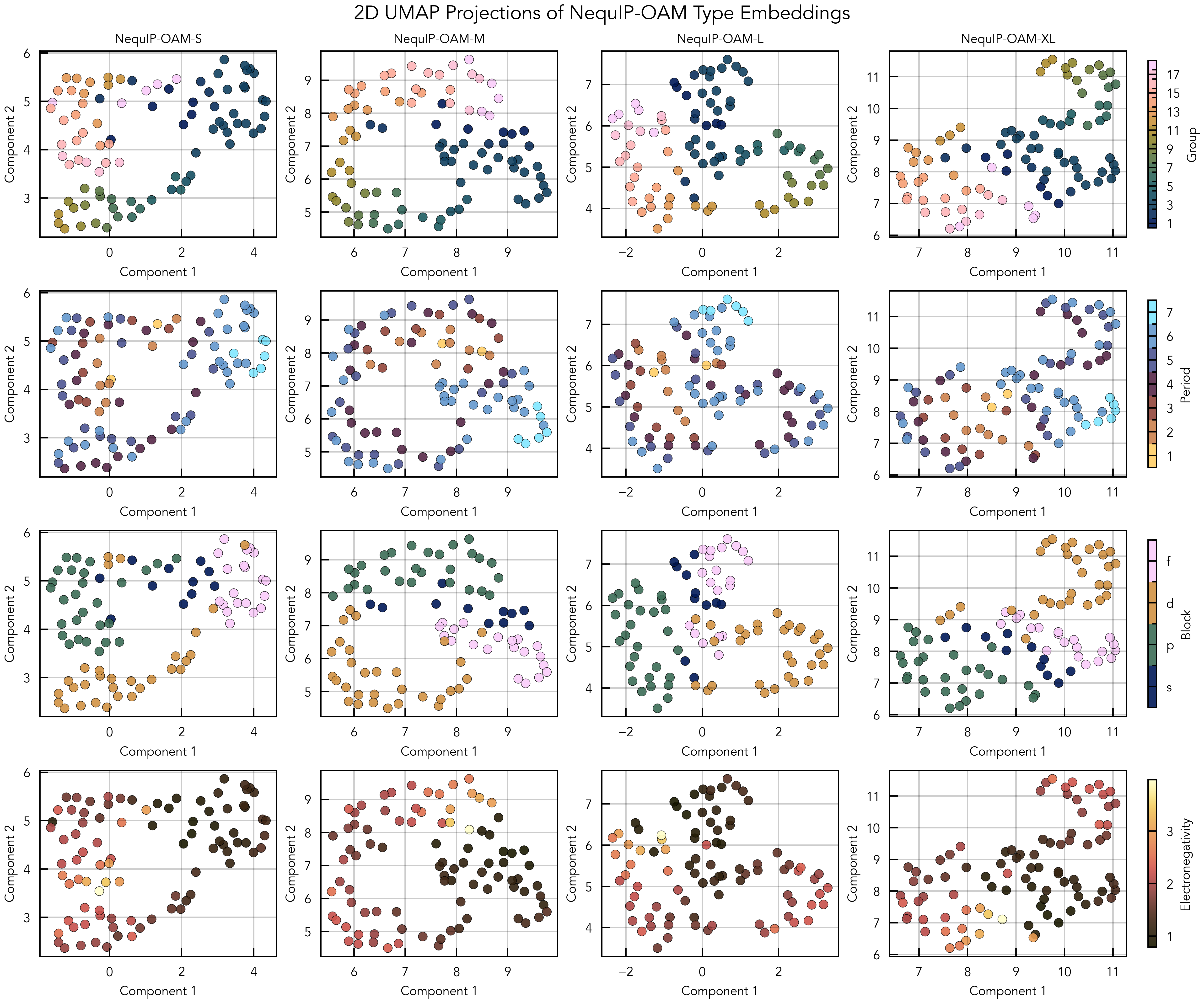}
\caption{\textbf{Uniform Manifold Approximation and Projection (UMAP) Dimensionality Reduction of Chemical Type Embeddings in NequIP-OAM Models}.
UMAP 2D projection plot of the chemical type embedding vectors learned by the NequIP-OAM models (S, M, L, XL; columns).
Scatter points are coloured by periodic group, period (row), periodic-table block, and Pauling electronegativity, in order from the top to bottom row. 
}
\label{sifig:umap_grid}
\end{figure}

\begin{figure}[htbp]
\centering
\includegraphics[width=\linewidth]{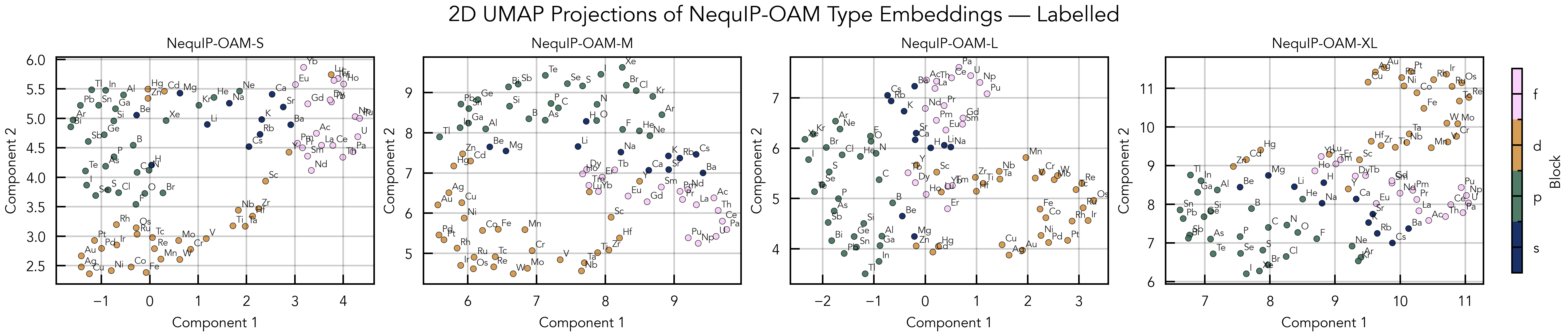}
\caption{\textbf{UMAP Dimensionality Reduction of Chemical Type Embeddings in NequIP-OAM Models -- Labelled}.
UMAP 2D projection plot of the chemical type embedding vectors learned by the NequIP-OAM models (S, M, L, XL; columns), labelled by the elemental identities.
Scatter points are coloured by periodic-table block.
}
\label{sifig:umap_labeled_all}
\end{figure}

\begin{figure}[htbp]
\centering
\includegraphics[width=\linewidth]{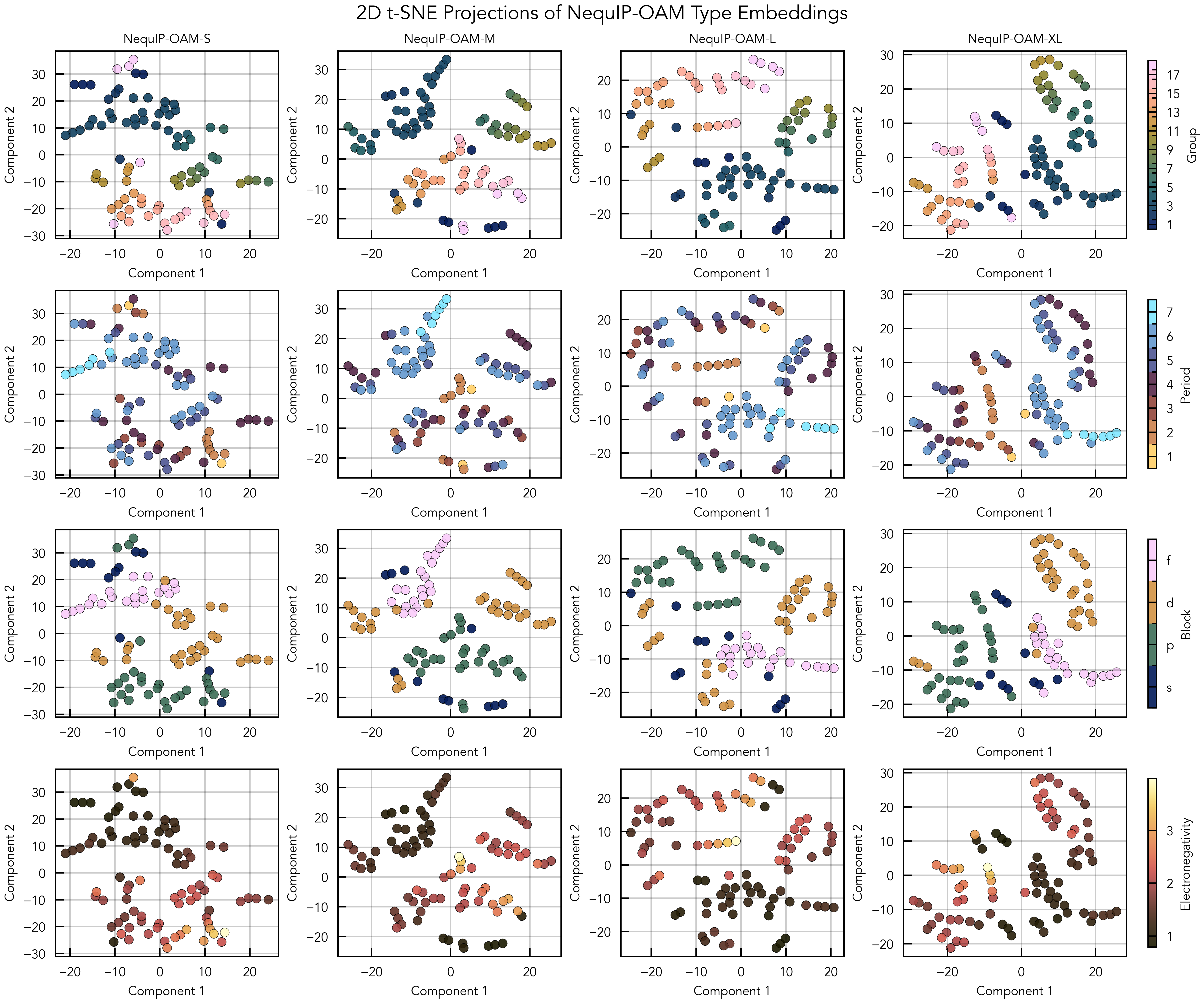}
\caption{\textbf{$t$-distributed Stochastic Neighbor Embedding ($t$-SNE) Dimensionality Reduction of Chemical Type Embeddings NequIP-OAM Models}.
$t$-SNE 2D projection plot of the chemical type embedding vectors learned by the NequIP-OAM models (S, M, L, XL; columns).
Scatter points are coloured by periodic group, period (row), periodic-table block, and Pauling electronegativity, in order from the top to bottom row. 
}
\label{sifig:tsne_grid}
\end{figure}

\begin{figure}[htbp]
\centering
\includegraphics[width=\linewidth]{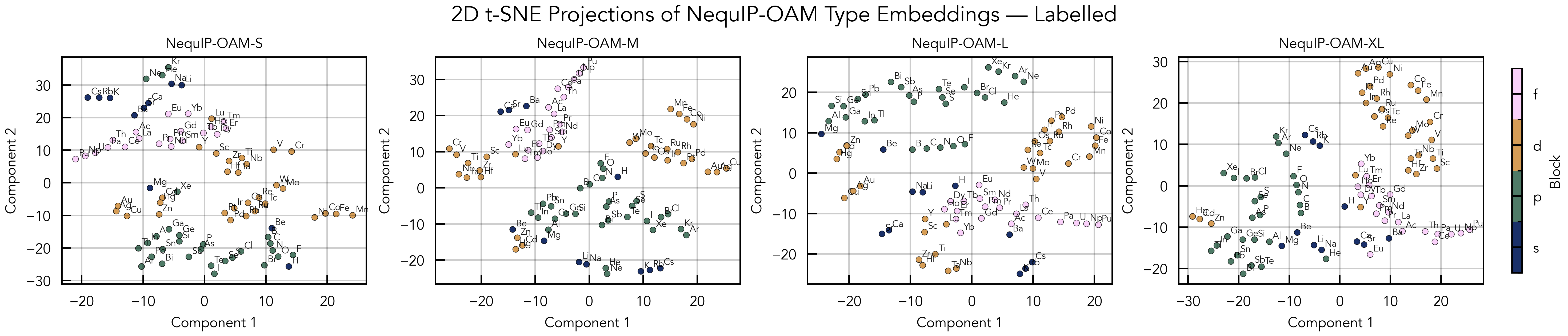}
\caption{\textbf{$t$-SNE Dimensionality Reduction of Chemical Type Embeddings in NequIP-OAM Models -- Labelled}.
$t$-SNE 2D projection plot of the chemical type embedding vectors learned by the NequIP-OAM models (S, M, L, XL; columns), labelled by the elemental identities.
Scatter points are coloured by periodic-table block.
}
\label{sifig:tsne_labeled_all}
\end{figure}

Clear chemical organisation is demonstrated by each dimensionality reduction analysis of the chemical type (elemental) embeddings across all NequIP-OAM models, shown in \cref{sifig:pca_grid,sifig:pca_labeled_all,sifig:umap_grid,sifig:umap_labeled_all,sifig:tsne_grid,sifig:tsne_labeled_all}.
This includes separation of periodic groups and periodic-table blocks ($s$, $p$, $d$, $f$), clear electronegativity gradients along reduced dimensions, and clustering of chemically-similar elements (e.g. coinage metals, halogens, lanthanides, noble gases \cref{sifig:pca_labeled_all,sifig:umap_labeled_all,sifig:tsne_labeled_all}) -- with progressively cleaner structure at larger model sizes. 
Noble gases and heavy alkali metals consistently appear as outliers under top-2 Principal Component Analysis (PCA) analysis.

For dimensionality reduction analysis, all type embeddings were standardised to have zero mean and unit variance (using \texttt{StandardScaler} from \texttt{scikit-learn}\cite{pedregosa_scikitlearn_2011}), prior to reduction. 
A fixed seed (\texttt{random\_state=7}) was used across all reductions for reproducibility.
The number of features for the type embedding layers (\texttt{type\_embed\_num\_features} in NequIP) in the NequIP-OAM models are 32 for S, M and XL, and 48 for L.
In \cref{sifig:pca_grid,sifig:pca_labeled_all,sifig:umap_grid,sifig:umap_labeled_all,sifig:tsne_grid,sifig:tsne_labeled_all} there are 89 data-points, corresponding to the 89 elements included in the OAM dataset (\cref{fig:per-element_errors,sifig:eSEN_per-element_errors}).
Uniform Manifold Approximation and Projection (UMAP) dimensionality reduction was performed using the \texttt{umap} package,\cite{mcinnes_umap_2018} with \texttt{n\_neighbors=15}, \texttt{min\_dist=0.1}, \texttt{metric="euclidean"} and \texttt{n\_components=2}.
For $t$-Distributed Stochastic Neighbor Embedding ($t$-SNE) dimensionality reduction, a perplexity value of 10 and PCA initialisation was used, via the \texttt{scikit-learn}\cite{pedregosa_scikitlearn_2011} package.

\FloatBarrier 
\putbib  
\end{bibunit}
\end{document}